\documentclass[11pt]{article}
\usepackage[a4paper,textwidth=180mm,top=16mm,bottom=16mm]{geometry}
\usepackage{amsmath,amsthm}
\usepackage{bm}
\usepackage{array,tabularx,float,needspace}
\newcolumntype{L}{>{\raggedright\arraybackslash}X}
\usepackage{newtxtext,newtxmath,graphicx,booktabs,microtype}
\usepackage[font=small,labelfont=bf]{caption}
\usepackage[hidelinks,pdfauthor={Shuoshuo Wang},pdftitle={Pachinko: Inferring founder-level outcomes of biological interventions from aggregate counts}]{hyperref}
\newtheorem{proposition}{Proposition}
\newcommand{\E}{\mathbb{E}}
\newcommand{\Pp}{\mathbb{P}}
\newcommand{\doop}{\operatorname{do}}
\title{Pachinko: Inferring founder-level outcomes of biological interventions from aggregate counts}
\author{\large Shuoshuo Wang\textsuperscript{1,2}\\[5pt]
\small \textsuperscript{1}Department of Pathology, Beth Israel Deaconess Medical Center,\\
\small Harvard Medical School, Boston, MA, USA\\
\small \textsuperscript{2}Cancer Program, Broad Institute of MIT and Harvard, Cambridge, MA, USA\\[3pt]
\small Correspondence: \href{mailto:wangshuo@broadinstitute.org}{wangshuo@broadinstitute.org}\\
\small ORCID: \href{https://orcid.org/0000-0002-8216-6164}{0000-0002-8216-6164}}
\date{}
\begin{document}
\maketitle
\begin{abstract}
Biological evolution is not merely movement through a landscape of possibilities; it is a process through which each successful transition may reshape the landscape of subsequent possibilities. Measuring these changes requires distinguishing outcomes of original founders from counts of their descendants. We develop a mathematical foundation for recovering founder outcome distributions from aggregate counts with calibrated references. One identity uses a reference proportional to latent founder intensity under conditional Poisson seeding; a second uses a reference proportional to actual founder number and permits arbitrary seeding distributions. Both allow outcomes within a founder's lineage to be dependent but require independence between founders and a common outcome distribution that does not change with exposure. Joint outcome probabilities can be recovered even when marginal event frequencies are unchanged. Across 5,800 synthetic two-arm experiments we examine finite-sample performance, alternative seeding, within-lineage selection, exposure-dependent response and reference mismatch. We distinguish information supplied by direct founder counts, reference channels and lineage tracking, and derive calibration sensitivity and a sharp bound on viable fate switching. Exact examples clarify why increased conditional competence need not increase founder-level success and why complete passive observations need not identify a response to challenge. The identities use classical generating-function and size-bias arguments. The contribution is a calibrated measurement framework for defined outcome laws, not a universal dynamical theory or empirical demonstration of a biological mechanism.

\end{abstract}

\section{What can a biological system become?}
Consider two cells that look identical under every available measurement. Both can divide, both express the same markers, and neither appears to be differentiating. A subsequent perturbation nevertheless permits one cell to generate a new fate while the other cannot. Did their histories change what they could become? Can an experiment distinguish this explanation from unequal growth, death or the limitations of the measurement itself?

The problem has a long scientific history. Waddington's epigenetic landscape gave developmental biology a visual language for constrained trajectories \cite{waddington}. Luria and Delbr\"uck used variation among bacterial cultures to distinguish alternative accounts of the origin of resistance \cite{luria}; fluctuation analysis remains a powerful illustration of why descendant counts cannot be equated with the number of original events \cite{rosche}. Blount, Borland and Lenski later used replay experiments to investigate historical contingency in the evolution of citrate utilization \cite{blount}. Modern trajectory methods such as Waddington-OT infer relationships among destructive population snapshots \cite{schiebinger}, while lineage tracing directly connects some antecedent states and later outcomes \cite{weinreb}. These traditions do not form a single linear history, and the idea that prior events matter is not new. They motivate a more specific question: can we measure a change in the future repertoire while accounting for the way cells branch and become observable?

These historical examples illustrate three different kinds of scientific inference. Fluctuation analysis uses the distribution of an observable, not simply its mean, to infer a hidden event process. Lineage tracing supplies a physical link between an ancestor and descendants; that link cannot be recovered from all population snapshots without further assumptions. A replay or intervention changes the experiment itself to reveal a dependency on history. Pachinko brings these questions together around one target: a founder's distribution of outcomes after a specified operation and challenge. It does not seek to reconstruct a unique ancestral history when that history is unavailable. The approach is complementary to multistage clonal expansion models, which explicitly connect mutation and population dynamics to disease incidence \cite{brouwer}. The contribution here concerns the observation design for a specified founder outcome, not a replacement for those dynamical models.

The key practical distinction is between counting target descendants and counting the original founders that produced them. A culture with one founder that produces ten target cells and a culture with ten founders producing one target cell each have identical aggregate target counts, but very different founder-level outcomes. The mathematical results below show what additional measurements can recover this difference, and when that recovery is impossible.

The name Pachinko refers to a familiar mechanical game in which a falling ball passes through a series of obstacles. The analogy becomes biological when passage can alter the obstacles encountered later. Cells differ from balls in essential respects: they divide, transmit some changes but not others, alter their surroundings and die. The analogy therefore guides the questions rather than defining a new mathematical class. An augmented stochastic state can encode these dynamics. Our objective is to infer particular biologically meaningful quantities, and to specify measurements that resolve ambiguity.

\section{From recurrence to a testable explanation}
Pachinko originated in cytogenomic questions about recurrent and sequential chromosome alterations. The present article deliberately separates mathematical foundations from disease-specific instantiations. A recurrent event can reflect its formation, acceptance, viable establishment, propagation or ascertainment. Observed enrichment alone does not determine which process changed.

\begin{figure}[!htbp]\centering
\includegraphics[width=\linewidth]{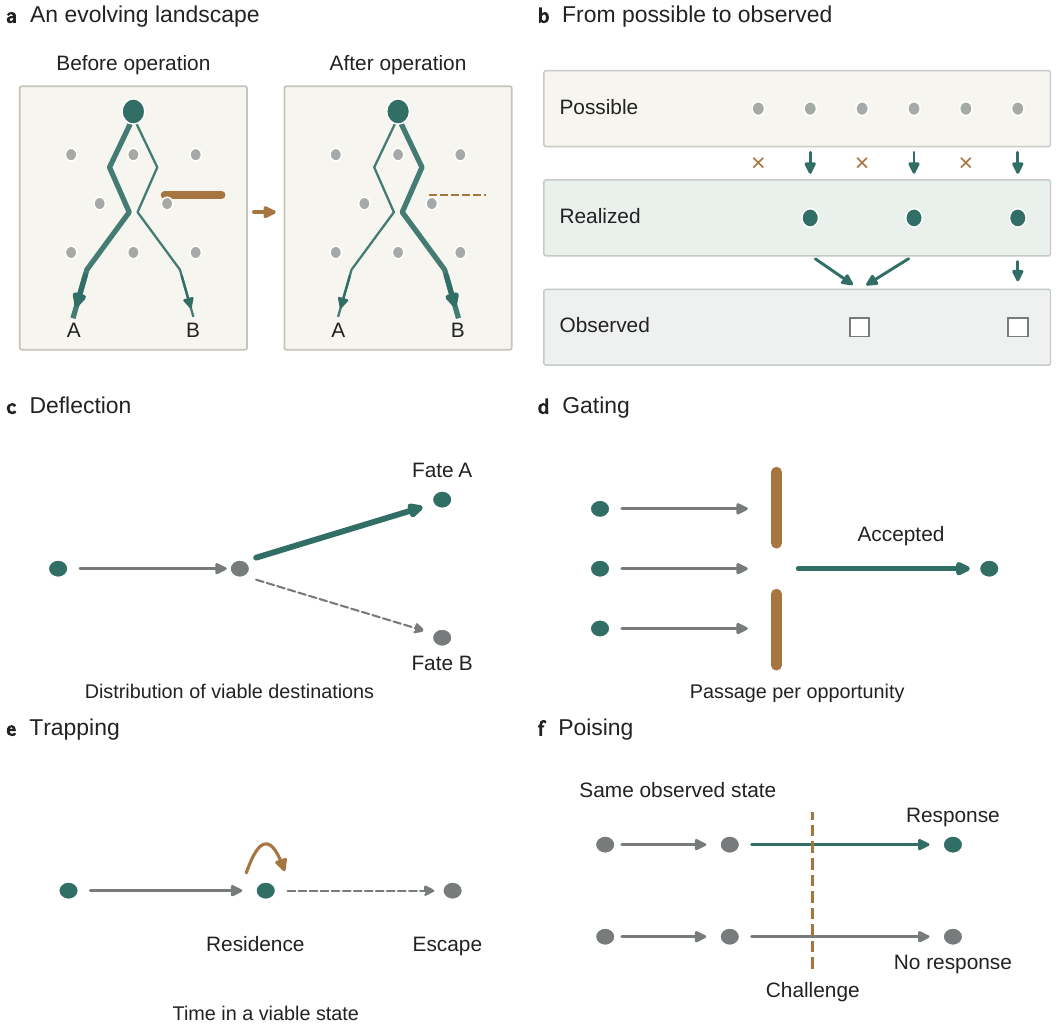}
\caption{Conceptual architecture of Pachinko. \textbf{a,} An antecedent operation can change the relative accessibility of two subsequent destinations, A and B. The obstacle and altered route represent a change in passage conditions; line widths are illustrative, not estimated transition probabilities. \textbf{b,} Possible histories (grey circles) are filtered by formation, survival and expansion into realized histories (green circles). Assay resolution can merge distinct realized histories into one observed category (squares). A missing observation does not establish biological impossibility. \textbf{c,} Deflection changes the distribution of viable destinations; distinguish it from differential death or expansion by measuring absolute founder outcomes. \textbf{d,} Gating changes accepted passage per opportunity; count proposals and rejected outcomes as well as successful passage. \textbf{e,} Trapping changes residence time or escape from a viable state; abundance alone does not establish retention. \textbf{f,} Poising is a difference in response to a specified subsequent challenge despite similar measured baseline states. Compare operation and sham histories under a matched challenge. These four functions are candidate experimental descriptions, not mutually exclusive mechanisms or an empirically validated classification.}\label{fig:overview}
\end{figure}

Three linked questions organize the investigation: which transitions are physically possible; which occur and leave viable descendants; and which are represented in the assay? Figure~\ref{fig:overview}a,b illustrates these distinctions. The same endpoint may admit several structural histories; different histories may share a measured phenotype. Conversely, a similar phenotype can conceal different functional dependencies. We use \emph{filtered recurrence} as a descriptive synthesis of these distinctions, not as proof that selection or any other particular filter caused recurrence. An observed zero is not proof of impossibility. For example, even if each of 100 independent founders has a one-percent chance of a target, the probability of no successes is $0.99^{100}$, about 37\%. The absence of a measured event constrains its probability only relative to the denominator, assay and experimental conditions.

\needspace{5\baselineskip}
A useful way to think about the ambiguity is to distinguish a history from its representation. If ten admissible histories lead to the same observed configuration, the observation alone cannot tell us which occurred. The number ten is not a biological probability: probabilities require a model of how histories are generated and retained. Even identical representations of past histories need not imply identical responses to a future intervention.

\section{Four ways to change the future}
An aperture is a name for a biological feature that may change which subsequent events occur, how long a state persists, or how a cell responds to a later challenge. It is not a literal genomic opening. It may involve chromosome architecture, gene regulation, differentiation, metabolism, stress tolerance or the environment. We distinguish four candidate operational functions (Figure~\ref{fig:overview}c--f and Table~\ref{tab:aperture}). They can coexist; calling a feature an aperture does not establish that it performs any of them.

\begin{table}[!htb]\centering\small
\caption{How to distinguish the four aperture functions experimentally.}\label{tab:aperture}
\begin{tabularx}{\linewidth}{@{}p{.85in}LL@{}}\toprule
Function & What would change? & What must be distinguished?\\\midrule
Deflection & Distribution across viable destinations & Differential death and growth can alter the same fractions. Measure absolute outcomes and competing loss.\\
Gating & Probability of accepted passage per opportunity & More divisions, more proposals or better detection can mimic increased tolerance.\\
Trapping & Residence time or escape from a viable state & Expansion, influx and delayed progression can mimic retention.\\
Poising & Response to a later specified challenge despite similar baseline behavior & Selection of pre-existing responsive cells or nonspecific growth can mimic acquired competence.\\\bottomrule
\end{tabularx}\end{table}

The term poising is used here operationally, rather than as a new theory of epigenetic priming. A factorial experiment with operation or sham, each with and without challenge, can quantify a response interaction. Baseline equivalence must be defined relative to measurements and a tolerance for meaningful differences; it is not established by a nonsignificant test. Trapping is not equivalent to death, and a productive gate may carry a cost. The four names are prompts for discriminating experiments, not mutually exclusive biological categories.

\subsection{A lower bound on fate redirection}
The distinction between deflection and selection has an elementary quantitative consequence. Suppose each arm starts with 100 eligible founders, and each founder has exactly one of two mutually exclusive viable terminal outcomes, A and B, or a nonviable or unresolved outcome N. In sham the expected counts are 50 A, 30 B and 20 N; after assignment to an operation they are 20 A, 60 B and 20 N. The gain of 30 B cannot be supplied entirely from the 20 founders in N. Even allowing rescue of every member of N, at least ten founders per hundred must have different viable destinations in any joint assignment of the two marginal fate distributions. If rescue is excluded, the minimum rises to thirty.

This is not tracking of ten individual cells. The argument belongs to the tradition of bounds over joint distributions with fixed marginals \cite{lu}. Selection bounds and principal stratification address related but different questions \cite{frangakis,zhang,lee}. It is a lower bound on a counterfactual coupling: a statement about all possible joint distributions consistent with the two observed marginals. It requires an original-founder denominator, a common target definition, mutually exclusive outcomes and comparable populations (for a causal interpretation, randomized assignment and the usual causal assumptions). It is not valid to substitute bulk transcript abundance or proportions among survivors. Appendix~\ref{app:coupling} provides the general formula and its sharpness argument.

\section{The quantity we want to measure}
\begin{figure}[!htbp]\centering
\includegraphics[width=\linewidth]{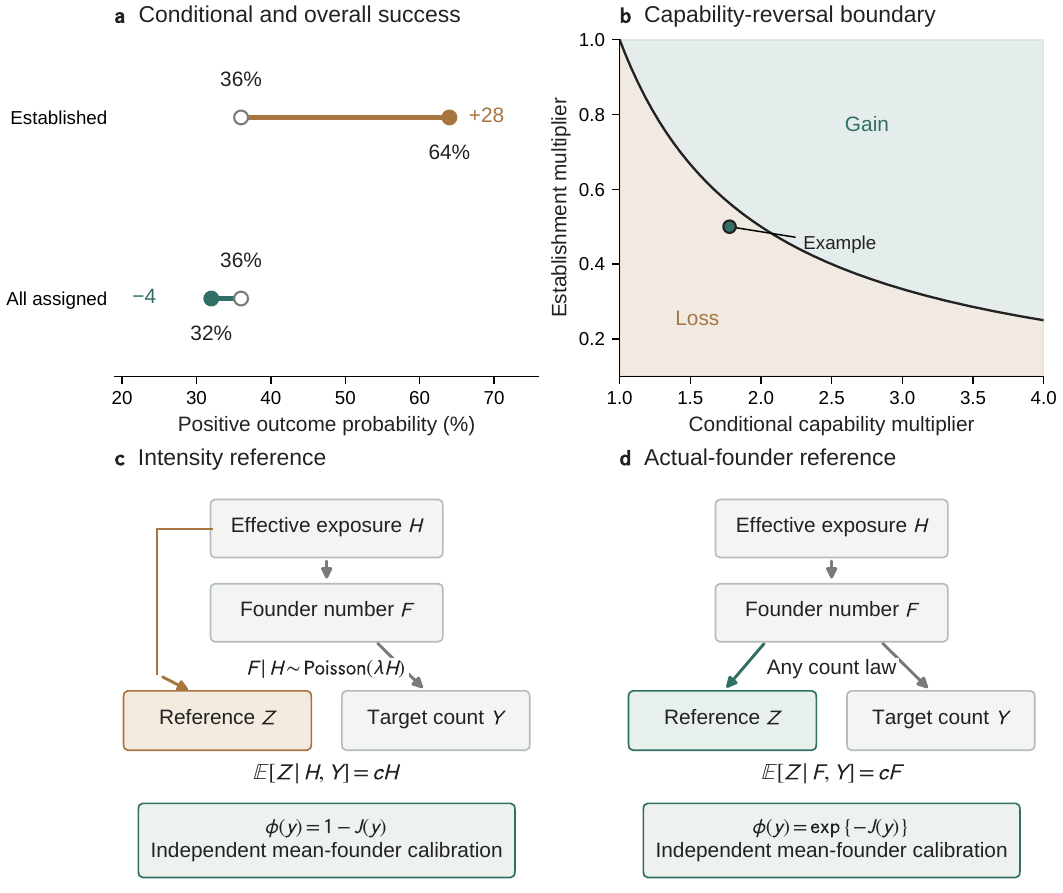}
\caption{From an original-founder denominator to calibrated aggregate measurement. \textbf{a,} Exact illustrative example: each established founder produces two independent daughters, each having a target probability of 0.20 in sham (open circle) or 0.40 after an operation (filled circle). All sham founders establish, whereas only half of operation-assigned founders establish. At least one daughter succeeds with probabilities 0.36 and 0.64 among established founders, but 0.36 and 0.32 among all originally assigned founders. The signed annotations indicate differences in percentage points. Overall success decreases by four percentage points, despite increased conditional competence. \textbf{b,} For establishment probability $a$ and conditional success probability $q$, overall success is $aq$ if failed establishment precludes the target. The boundary $a_e/a_0=q_0/q_e$ separates gain (green) from loss (ochre); the marked example has establishment multiplier 0.5 and conditional multiplier $0.64/0.36$. Feasible multipliers depend on baseline probabilities. \textbf{c,} An intensity reference $Z$ tracks effective exposure $H$, with founder count $F\mid H\sim\mathrm{Poisson}(\lambda H)$ and $\E[Z\mid H,Y]=cH$. The target count $Y$ sums founder outcomes, and $\lambda,c>0$. \textbf{d,} An actual-founder reference instead satisfies $\E[Z\mid F,Y]=cF$ and permits any founder-count distribution with finite positive mean. Both designs require independent calibration of mean founder number $m=\E[F]$ to identify the unconditional outcome law $\phi(y)=\E[y^V]$ for the outcome count $V$ of one founder. The shared integral $J(y)$ is defined in equation~\eqref{eq:Jmain}; its linear and exponential transformations are valid under different physical measurement assumptions. Panels a and b are exact calculations; panels c and d are proposed measurement designs.}\label{fig:measurement}
\end{figure}

Imagine two experiments, each beginning with 100 eligible founder cells. A founder is an original cell whose subsequent outcomes are of interest, whether or not that cell survives or produces a target outcome. In the sham arm every founder establishes and produces two daughters. Each daughter has a 20\% chance of the target outcome. The chance that at least one daughter succeeds is 36\%, since both fail with probability $0.8\times0.8$. In the operation arm the daughter probability doubles to 40\%, but only half the original founders establish. Among established founders the probability of a target is 64\%; among the original founders it is only 32\%. Thus, in this example, the operation improves conditional competence while reducing overall capability by four percentage points. The values are expectations under an illustrative model, not actual cell counts (Figure~\ref{fig:measurement}a).

The general criterion is equally simple. Write $a_e$ for establishment probability and $q_e$ for the conditional probability of a positive outcome after establishment. If failure of establishment precludes a positive outcome, then $C_e=a_eq_e$. When $q_e>q_0>0$ and $a_0>0$, improved conditional capability reduces overall capability exactly when $a_e/a_0<q_0/q_e$. The boundary is a hyperbola in the two multipliers (Figure~\ref{fig:measurement}b). This is a statement about a composite outcome defined as zero for failed establishment, not a survivor-average causal effect with an outcome undefined after death. The latter raises the distinct principal-stratification problem \cite{frangakis,zhang}. Appendix~\ref{app:branch} gives extensions to random daughter counts and correlated outcomes.

More formally, let $z$ denote the starting configuration, $e$ the assigned operation, $u$ a specified subsequent challenge or sequence, $B$ the target and $T$ the horizon. Let $E_B(T)$ be the explicitly defined outcome, for example at least one viable target descendant at time $T$. We write
\begin{align}
 C_T(z,e,u,B) &:= \Pp\bigl(E_B(T)\mid z,\doop(e),\doop(u)\bigr),\label{eq:capability}\\
 \Delta_e C_T &:= C_T(z,e,u,B)-C_T(z,0,u,B).\label{eq:capability_delta}
\end{align}
The second term is the sham arm. The denominator is the number of original founders. It includes those that die, fail to establish or fail to acquire the assigned change, rather than only descendants or successful edits. Operation assignment is not synonymous with successful acquisition. A sequence of challenges can be fixed in advance or governed by a specified policy. The family of outcome laws over these choices is the intervention-conditioned repertoire. Its entries need not share a single scalar notion of fitness. A change in probability is not necessarily a change in strict mathematical support.

The operation can be a structural rearrangement, a reversible regulatory perturbation or an environmental exposure. The symbol $z$ describes the specified pretreatment background, not a requirement that all microscopic variables be known. If assignment does not always induce acquisition, the contrast in equation~\eqref{eq:capability_delta} is an assignment effect. Estimation of an effect of acquisition itself would require an additional intervention or identification strategy. Similarly, a dynamic challenge policy $u$ is understood as an intervention rule fixed in advance, including the measurements on which it may depend. The definition neither assumes nor requires a low-dimensional hidden state. For a repertoire, compare a prespecified collection of these contrasts across targets, challenges and horizons; do not treat the sum of probabilities of overlapping events as a probability of reaching any one of them.

Counting whether a target ever appeared, whether it survived until sampling, how many cells it produced, and whether the assay detected it answers different questions. If no lineage information exists, a cell proportion alone does not yield the founder denominator. Some quantities cannot be uniquely determined from the available measurements. Recognizing this limitation is part of the analysis; it should not be concealed by replacing the question with an easier one.

\section{Recovering outcomes without reconstructing every lineage}
Suppose many independent cultures contain unknown numbers of founders. A larger culture may produce more of the target simply because more cells started there. If we can introduce a reference measurement that increases proportionally with founder exposure, and that does not otherwise respond to the target biology, it can help separate these effects (Figure~\ref{fig:measurement}c,d). A hypothetical design is independent neutral reference founders introduced by the same dilution process; their suitability must be tested, not assumed. A read-depth control alone is not necessarily such a reference.

We use three quantities: $Y$, the aggregate number of counted outcomes in a culture; $Z$, its reference count; and $m$, the independently calibrated mean number of original eligible founders. An optional primary count $X$ records detected primary founders, not primary descendant cells. Within a founder, branching, survival and secondary outcomes may be dependent. Across founders the marks must be independent with a common distribution invariant to the seeding variable. Two reference architectures are distinguished.

In the intensity design, $F\mid H\sim\operatorname{Poisson}(\lambda H)$ and $\E[Z\mid H,Y]=cH$, where $H$ is the effective exposure and $c>0$. The full distribution of $H$ is not needed. Mixed-Poisson seeding is allowed when the reference tracks the entire latent intensity. In the actual-founder design, $F$ can have any nonnegative integer distribution with finite positive mean, but the stronger and different physical measurement is $\E[Z\mid F,Y]=cF$. An independent pre-intervention per-founder signal could in principle serve this role; neither an arbitrary sequencing control nor a second nominal dilution automatically qualifies.

The probability-generating function, $\phi(y)=\E[y^V]$, encodes the outcome count $V$ of one founder. For example, if a founder can produce zero, one or two counted outcomes, with probabilities $p_0,p_1,p_2$, then $\phi(y)=p_0+p_1y+p_2y^2$. Evaluating the function at $y=0$ gives the probability of no counted target. Evaluating it at $y=1$ gives one, because all probabilities sum to one. Define the measured integral
\begin{equation}\label{eq:Jmain}
 J(y)=\frac{\E[Z]}{m}\int_y^1\frac{\E[Ys^{Y-1}]}{\E[Zs^Y]}\,\mathrm{d}s.
\end{equation}
The two identities are
\begin{align}
 \phi(y)&=1-J(y) &&\text{(intensity reference)},\label{eq:mainlin}\\
 \phi(y)&=\exp\{-J(y)\} &&\text{(actual-founder reference)}.\label{eq:mainlog}
\end{align}
The corresponding positive-outcome estimates are $J(0)$ and $1-\exp\{-J(0)\}$. These formulas use the same numerical integral but not the same experimental assumptions. They cannot be interchanged on a given reference without calibration. In either design a separate mean-founder measurement anchors the unconditional scale. In the intensity design the primary count and reference identify the detected-founder conditional law without that calibration. The proofs, boundary conventions and joint extensions are in Appendix~\ref{app:recovery}.

\subsection{Why the reference works}
Imagine two sets of cultures with the same founder biology but different numbers of initial cells. The probability of an empty target channel is smaller in the more densely seeded set. If seeding is unmeasured, the distribution of aggregate target counts blends variation in founder number with variation in individual outcomes. Our reference design is intended to sample the first process independently of the second. In the recovery equation, the reference is not simply a divisor for the mean count: target counts weight both measurements at each point of a generating function. The two weighted measurements contain a common mathematical factor representing unknown seeding variation. Dividing cancels this factor. Integrating the resulting derivative, with the requirement that a probability distribution sums to one, reconstructs the founder law. The mean founder calibration sets the absolute scale. The cancellation is linear in the founder generating function for a conditional Poisson intensity reference; it is logarithmic for a reference proportional to realized founder count. Both are size-weighted generating-function identities. The argument works for a multidimensional nuisance vector if the corresponding conditional mean restriction holds. Appendix~\ref{app:calibration} explains what happens when it is not.

This is related to ecological $N$-mixture models, in which replicated counts are used to infer latent abundance; their identifiability and sensitivity to unverified measurement assumptions are well documented \cite{dennis,barker}. The Poisson marking identities employed here are classical. Our use of them is specifically to recover a defined distribution of biological outcomes and to make founder calibration and experimental design explicit. No claim is made that aggregate counts suffice for all biological questions, nor that reference normalization has no precedent in other fields.

\subsection{Joint outcomes: beyond a single target}
Consider two distinct outcomes A and B, measured in the same experimental cultures. Each is present in 35\% of original founder lineages in both arms. Nevertheless, the sham probability of both is 10\%, whereas after the operation it is 25\%. The probabilities of either event are 60\% and 45\%, respectively. Thus every marginal comparison is unchanged, yet the joint pattern changes substantially. The joint recovery experiment is described below.

This dependency can be recovered from aggregate target counts without knowing which founder produced each counted outcome. Apply the appropriate scalar recovery formula to $Y_A$, $Y_B$ and $Y_A+Y_B$. The third positive-outcome probability is that of the union, not the sum of individual probabilities. Inclusion--exclusion then gives
\begin{equation}\label{eq:intersection}
 p_{A\cap B}=p_A+p_B-p_{A\cup B}.
\end{equation}
The construction extends to binary outcome patterns across a finite panel and to a multivariate count generating function (Appendix~\ref{app:recovery}). Importantly, ``both'' means that both outcomes occurred within the same founder lineage; it does not imply their co-residence in a single descendant, their order of acquisition or a joint counterfactual across treatment arms. Full distribution identification at the population level is distinct from stable finite-sample reconstruction of all coefficients.

\section{Why more observations can strengthen a wrong conclusion}
\begin{figure}[!htbp]\centering
\includegraphics[width=\linewidth]{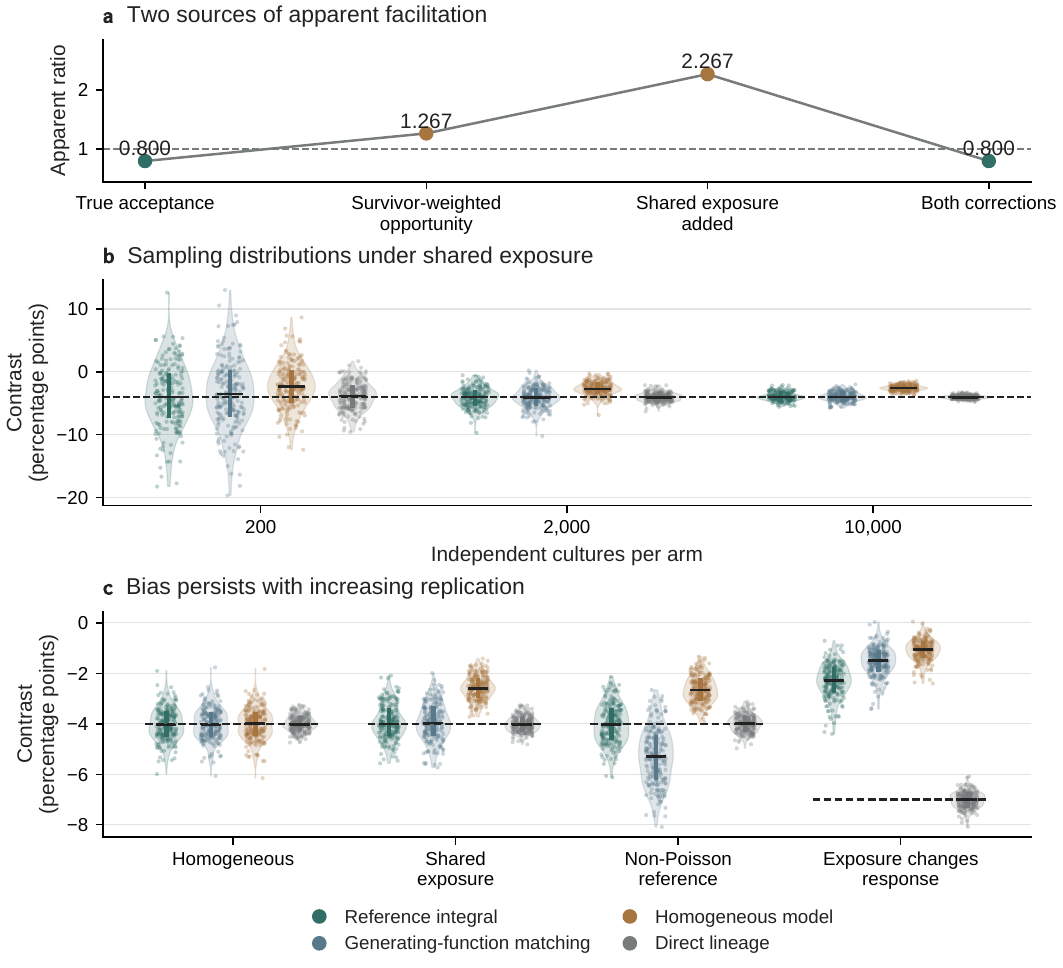}
\caption{Branching and exposure distortions, and finite-sample recovery. \textbf{a,} Exact critical birth--death example with birth and death rates one, terminal time one, immortal secondary targets, and survival-conditioned primary detection. A true acceptance ratio $4/5$ becomes $19/15$ after the survivor-conditioned opportunity multiplier $19/12$; common exposure variation with equiprobable $H=0.5$ or $1.5$ and mean five retained founders adds one to the naive ratio, giving $34/15$. Correcting both under their respective calibrations restores $4/5$. The horizontal dashed line marks neutrality, not a statistical threshold. \textbf{b,} Synthetic operation-minus-sham contrasts at 200, 2,000 and 10,000 independent cultures per arm under shared exposure $H=0.5$ or $1.5$ with equal probability, mean five founders and mean reference count eight. The two-daughter outcome model in Figure~\ref{fig:measurement}a has truth $-4$ percentage points. \textbf{c,} At 10,000 cultures per arm, the four scenarios use homogeneous exposure, shared exposure with a Poisson reference, shared exposure with a reference $Z=2\mathrm{Poisson}(4H)$, or exposure-dependent daughter success probabilities. The last scenario has founder-weighted truth $-7$ percentage points; all others have truth $-4$. Dashed black lines show the corresponding truths. The integral, generating-function matching and homogeneous estimators use aggregate measurements (the homogeneous estimator ignores exposure); the direct-lineage estimator uses true founder and positive-founder counts. Each violin is a kernel density of 200 independent simulated two-arm estimates (Scott bandwidth); jittered points are individual estimates, the colored segment spans the interquartile range and the black mark is the median. Jitter changes only categorical position. Data: retained simulation arrays and numerical summaries from \texttt{benchmark.py}; no empirical data.}\label{fig:performance}
\end{figure}

There are two distinct sources of apparent facilitation. A surviving clone has often accumulated more living-cell time than a random original founder. In one critical birth--death example with unit birth and death rates, survival-gated primary detection and immortal targets, the expected opportunity among surviving founders is $19/12$ times the unit reference opportunity. A true acceptance ratio of 0.8 can then appear to be 1.267. Variation in founder exposure across cultures can add another distortion: with the stated synthetic parameters it raises the naive ratio to 2.267. The two corrections are not interchangeable (Figure~\ref{fig:performance}a). A valid reference removes common exposure variation; a separate calibration of biological opportunity is needed to interpret a tolerance ratio. The occupation factor is not identified by the aggregate counts used here. Time-resolved lineage measurements or a separately validated living-cell-time proxy would be necessary to estimate it. Appendix~\ref{app:branch} derives both effects and states the necessary common-factor assumptions.

\section{Does the recovery procedure work?}
The first simulation compares the intensity-reference integral with an alternative generating-function matching estimator in a deliberately simple branching example. The actual-founder reference is evaluated separately below. Each arm has $N$ independent replicate cultures. Given exposure $H$, the number of assigned founders is Poisson with mean $5H$. In sham, each founder establishes and generates two daughters, each independently producing a counted target event with probability $0.2$. In the operation arm, establishment probability is $0.5$ and daughter probability is $0.4$. Thus the true founder-level outcome probabilities are $0.36$ and $0.32$, giving a contrast of $-0.04$. Failed establishment contributes zero. The mean number of original founders is known. A reference has conditional mean $8H$. No lineage label is used by either reference estimator.

We generated four scenarios: homogeneous $H=1$; equiprobable $H\in\{0.5,1.5\}$ with Poisson reference; the same exposure with $Z=2\operatorname{Poisson}(4H)$; and a violation in which daughter probability itself is multiplied by $H$. In the last case the founder-weighted target contrast is $-0.07$. In each setting and sample size ($N=200,2000,10000$ per arm), 200 independent Monte Carlo experiments were generated, for 2,400 two-arm experiments. Poisson thinning over the three possible mark categories gives an exact discrete-branching simulator. This benchmark is a synthetic finite-generation experiment, not a fitted biological cohort.

The integral estimator used empirical numerator and denominator and fixed 24-point Gauss--Legendre quadrature. The generating-function matching estimator used a one-dimensional root of the empirical reference generating function at the observed zero-event probability. The homogeneous comparator deliberately ignores exposure variation; it is an internal stress-test baseline, not a proxy for all established fluctuation or latent-count methods. A direct-lineage estimator uses the true simulated founder counts and positive-founder counts; it has access to more information and is not a fair matched-information competitor. Uncertainty used culture-level influence-function standard errors, combined across independent arms. Out-of-range values were not clipped. All 200 contrasts per method and scenario/sample-size setting were finite with finite standard errors; no estimated arm probability fell outside $[0,1]$ in these simulations. The complete output includes both failure counters and coverage over all planned repetitions. This is an initial finite-sample assessment, not a proof of uniform coverage.

\begin{table}[!htb]\centering\small
\caption{Synthetic contrasts under shared exposure. All values except coverage are percentage points; truth is $-4$ percentage points. The 95\% interval coverage is the empirical fraction covering that truth. 200 experiments per row.}\label{tab:benchmark}
\begin{tabular}{r l r r r}\toprule
Cultures per arm & Method & Mean contrast & Root mean squared error & Coverage (\%)\\\midrule
200 & Integral & -3.91 & 4.92 & 95.0\\
 & Generating-function matching & -3.53 & 5.46 & 95.0\\
 & Homogeneous & -2.30 & 3.93 & 92.5\\
 & Direct lineage & -3.81 & 2.14 & 95.5\\
2000 & Integral & -4.10 & 1.50 & 95.5\\
 & Generating-function matching & -4.09 & 1.57 & 96.0\\
 & Homogeneous & -2.67 & 1.69 & 74.5\\
 & Direct lineage & -4.03 & 0.69 & 95.5\\
10000 & Integral & -3.93 & 0.68 & 95.5\\
 & Generating-function matching & -3.94 & 0.73 & 94.5\\
 & Homogeneous & -2.57 & 1.50 & 17.5\\
 & Direct lineage & -4.01 & 0.30 & 94.0\\\bottomrule
\end{tabular}\end{table}

Under shared exposure with the stated calibration and reference law, both reference estimators recover the contrast at large sample size (Table~\ref{tab:benchmark}, Figure~\ref{fig:performance}b,c). At 200 cultures per arm the homogeneous comparator has smaller root mean squared error than either reference estimator despite bias; the benefit of model-correct reference normalization is not uniform in sample size. At 10,000 cultures per arm the homogeneous estimator has coverage of only 17.5\%, versus 95.5\% for the integral. Under the non-Poisson reference scenario the integral estimate is $-4.00$ percentage points (root mean squared error $0.76$ percentage points, coverage $93\%$), whereas generating function matching is biased ($-5.30$ percentage points; root mean squared error $1.70$ percentage points, coverage $71\%$). This comparison isolates the importance of requiring only the conditional mean reference law. Under the tested multiplicative exposure-dependent mark biology, the true contrast is $-7$ percentage points and the integral estimator gives $-2.30$ percentage points, with no interval covering the target at 10,000 cultures per arm. Increasing replication does not repair this model failure. The full table, including small-sample performance and numbers of valid estimates, is supplied with the code.

The practical lesson is not that reference normalization can recover a mechanistic effect in arbitrary heterogeneous cultures. It is that variation in founder number and variation in founder response are different biological processes, with different inferential consequences. Reference calibration needs to be established in a real experimental system. More informative parametric or semiparametric likelihoods, independent calibration of founder numbers and stronger biological baselines may improve efficiency and robustness.

\subsection{Stress testing selection and the reference channel}
The preceding example deliberately separates passage from establishment. To assess sensitivity to two biologically important departures, we carried out a second, separately seeded benchmark of 1,800 two-arm experiments. We introduced shared propensity among two daughters and subsequent target-dependent survival. In the selection model, an established founder has daughter acquisition probability $q-0.1$ or $q+0.1$ with equal probability. Conditional on this propensity, daughter events are independent; marginally they are correlated. Each target-bearing daughter yields a binomial number of terminal target descendants from two opportunities, with survival probability 0.85. Non-target daughters survive with probability 0.35 per grandchild opportunity and contribute no counted target; their counts are marginalized. The acquisition probabilities and establishment rates before this additional generation are the same as in the first benchmark. There is no competition between founders. This changes the distribution of marks without making that distribution exposure-dependent.

In a separate stress test the target exposure is $H\in\{0.5,1.5\}$ with equal probabilities, as before. The reference exposure $H_Z$ equals $H$ with probability $(1+\rho)/2$ and $2-H$ otherwise. Thus $\E[H_Z\mid H]=1+\rho(H-1)$, with $\rho=1$ representing exact matching and $\rho=0$ independent exposures. Conditional on $H_Z$, the reference count is Poisson with mean $8H_Z$. The six scenarios combine matching, mismatch in one or both arms, and selection. Each is evaluated at 200, 2,000 and 10,000 cultures per arm, with 100 independent simulations per setting. The exact finite mark probabilities were used for Poisson thinning to generate aggregate data; no real clonal samples are implied.

\begin{figure}[!htbp]\centering
\includegraphics[width=\linewidth]{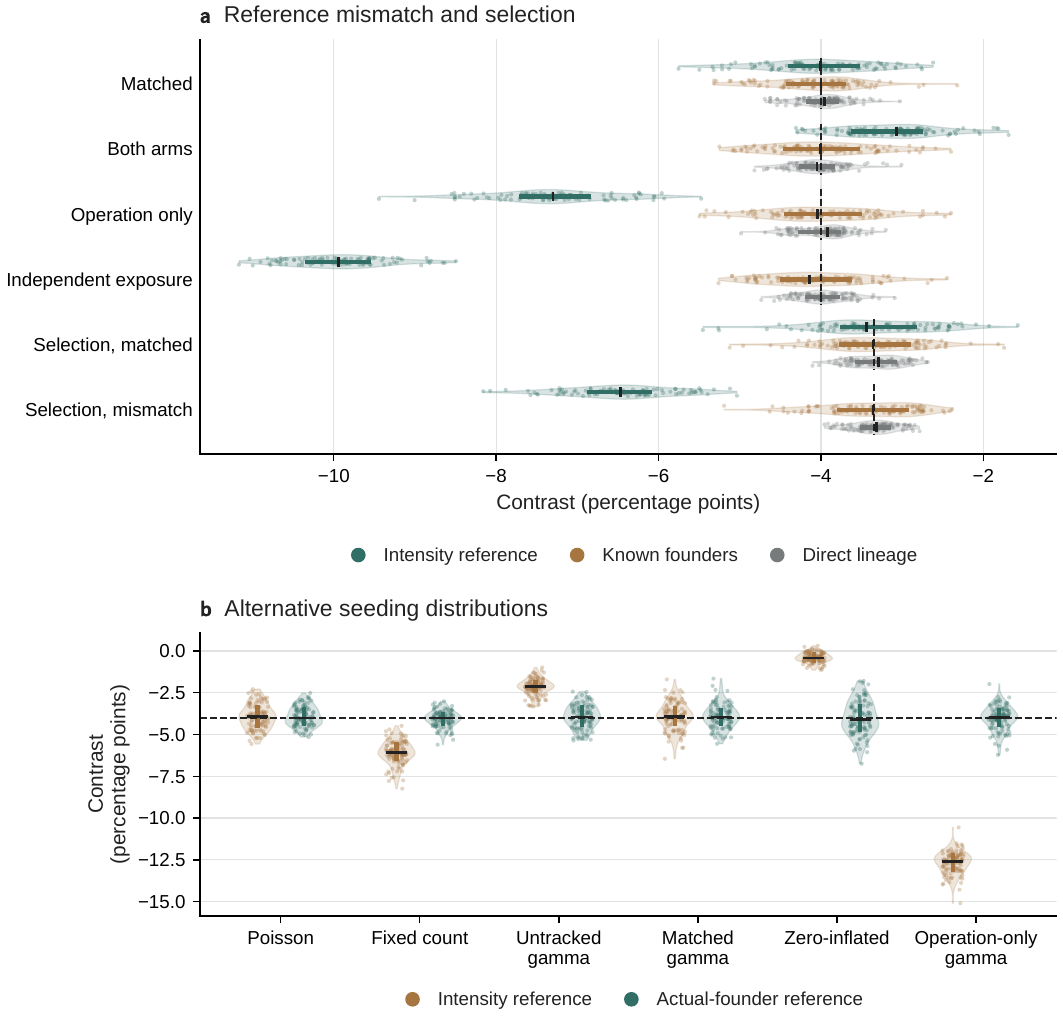}
\caption{Robustness to reference mismatch, within-lineage selection and non-Poisson seeding. \textbf{a,} Reference exposure tracks the target exposure exactly (matched), is partially mismatched with correlation 0.5 in both arms or only in the operation arm, or is independent in the operation arm. The final two rows include shared daughter propensity ($q\pm0.1$, with equal probability) and two terminal opportunities per daughter, with target-bearing and non-target survival probabilities 0.85 and 0.35. The selection model is matched or partially mismatched in the operation arm. The true contrast is $-4$ percentage points in the first four rows and $-3.344$ in the selection rows. The three designs use the intensity reference, known founder counts, or direct lineage information; the latter two observe additional quantities. \textbf{b,} Six seeding regimes at mean five founders: conditional Poisson; exactly five founders; an untracked gamma multiplier (shape two, scale one-half); a gamma multiplier fully tracked by the intensity reference; untracked zero inflation (multiplier zero or two with equal probability); and extra untracked gamma variation confined to the operation arm. The intensity reference tracks the declared exposure with mean eight; the actual-founder reference has conditional Poisson mean $1.6F$. The true contrast is $-4$ percentage points throughout. Each violin and its jittered estimates summarize 100 independent two-arm experiments with 10,000 cultures per arm. Colored segments indicate interquartile ranges, black marks medians, and dashed black lines the scenario-specific truth. The reference designs use different physical measurements, not identical inputs. The root mean squared errors at both sample sizes are provided in Table~\ref{tab:seedingprecision}. Data: retained arrays from \texttt{robustness.py} and \texttt{seeding.py}.}\label{fig:robustness}
\end{figure}

With a matched reference, the selection scenario has true contrast $-3.344$ percentage points and mean recovered contrast $-3.329$ at 10,000 cultures per arm (root mean squared error 0.711 percentage points; nominal 95\% coverage 94\%). Thus within-founder dependence and selection do not invalidate the recovery identity. This is a consequence of the theorem's arbitrary joint mark law, not evidence that selection in real tissues is independent across founders. The selection model is finite-generation and does not model niche competition.

With partial mismatch in the operation arm alone ($\rho=0.5$), the original contrast of $-4$ percentage points becomes $-7.30$ percentage points on average. With independent reference exposure in that arm it becomes $-9.93$ percentage points. In both cases none of the 100 nominal intervals covers the true effect at 10,000 cultures per arm. The same mismatch in both arms attenuates the contrast to $-3.14$ percentage points, with 69\% coverage. These are examples of differential measurement bias, not universal functions of correlation. Correlation alone cannot calibrate the reference. The known-founder estimator, which observes the original number of founders in each culture, gives $-4.01$ percentage points under partial operation-arm mismatch (root mean squared error 0.66 percentage points); direct lineage accounting gives $-3.99$ percentage points. Figure~\ref{fig:robustness}a shows these differences. All 100 repetitions at the largest sample size in these six scenarios gave finite estimates and standard errors, with no out-of-range arm estimates. The full results report failures across all sample sizes.

\subsection{Alternative seeding and joint outcome recovery}
To isolate the founder-number assumption, we ran a third suite of 1,200 two-arm experiments. Each setting used 100 repetitions at 2,000 and 10,000 cultures per arm, mean founder number five and the same two-daughter mark laws as the original example. The six seeding regimes were conditional Poisson, fixed five founders, gamma-mixed Poisson with an untracked multiplier, gamma-mixed Poisson with a fully matched intensity reference, zero-inflated Poisson with untracked zero inflation, and gamma mixing confined to the operation arm. Gamma multipliers had shape two and scale one-half. In the zero-inflated case the intensity was multiplied by zero or two with equal probability. Except for fixed seeding, the initial $H$ was equiprobably 0.5 or 1.5. The intensity reference had mean $8H$ (or eight times the full normalized intensity in the matched mixture); the actual-founder reference was Poisson with conditional mean $1.6F$. The known-founder likelihood and direct-lineage estimator use progressively more physical information.

At 10,000 cultures per arm, the intensity estimator returned mean contrasts of $-6.10$, $-2.16$, $-0.39$ and $-12.69$ percentage points under fixed, untracked gamma, untracked zero inflation and operation-only dispersion, respectively, despite the common truth $-4$ percentage points. The actual-founder estimator recovered means of $-4.07$, $-3.94$, $-4.02$ and $-4.02$ percentage points. Its nominal coverage across all six settings was 94--98\%. The intensity estimator also recovered the contrast in the fully matched mixed-Poisson design. Figure~\ref{fig:robustness}b shows the sampling distributions. The key distinction is whether the reference matches the seeding variable required by the identity, not whether marginal founder counts are overdispersed. The two reference designs differ in information; this is not an algorithm competition under identical observations.

\begin{figure}[!htbp]\centering
\includegraphics[width=\linewidth]{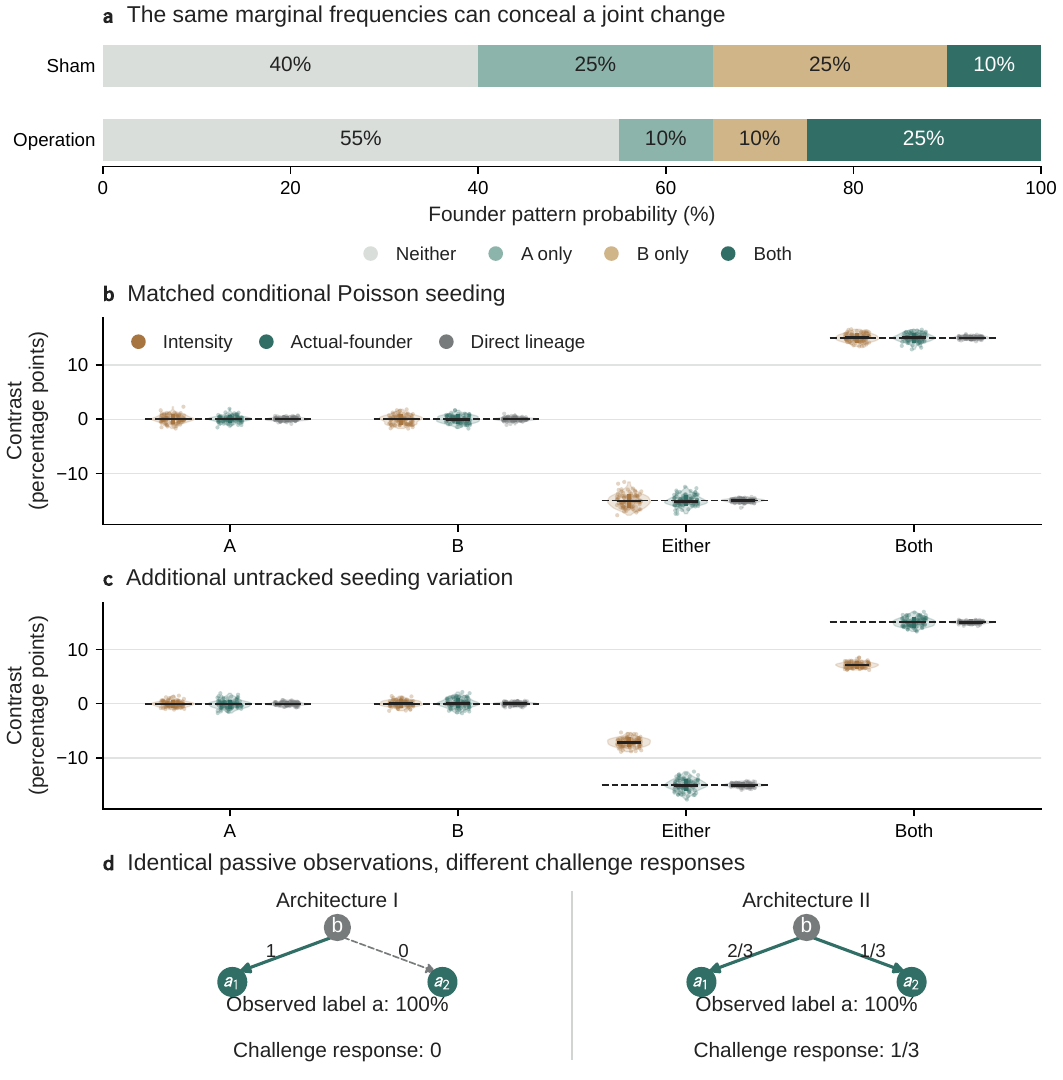}
\caption{Changes in joint founder outcomes and hidden challenge competence. \textbf{a,} Exact synthetic founder patterns (neither, A only, B only, both) are $(0.40,0.25,0.25,0.10)$ in sham and $(0.55,0.10,0.10,0.25)$ after operation. The probability of each individual event remains 0.35, whereas joint occurrence increases by 15 percentage points and occurrence of either decreases by 15. Both outcomes can occur within one founder lineage without being present in the same descendant cell. \textbf{b,c,} Recovery of the four contrasts using aggregate counts of A, B and their sum, under matched conditional Poisson seeding (b) or an additional untracked gamma multiplier (c; shape two, scale one-half). The intensity reference has mean $8H$, and the actual-founder reference has conditional mean $1.6F$. The direct-lineage estimator observes the true founder patterns. Each violin displays 100 independent two-arm estimates at 10,000 cultures per arm; points are individual estimates, segments interquartile ranges and black marks medians. Dashed black segments denote the exact contrast for each event. Data: retained arrays from \texttt{joint\_benchmark.py}. \textbf{d,} Two members of an exact four-state Markov family have the same complete baseline observation law. After the first transition, hidden states $a_1$ and $a_2$ share measured label $a$; the probability of entering $a_2$ is zero in architecture I and one-third in architecture II. A separately calibrated challenge responding only in $a_2$ therefore produces different response probabilities despite identical passive observations. Later transition probabilities also differ, preserving every baseline word probability. The full matrices and their similarity transformation are given in Appendix~\ref{app:challenge}; this panel shows only the first transition and challenge, not the entire transition graph. None of the panels represents empirical biology.}\label{fig:joint}
\end{figure}

A fourth suite tests the joint outcome example in equation~\eqref{eq:intersection}. The sham mark probabilities (neither, A only, B only, both) are $(0.40,0.25,0.25,0.10)$; operation probabilities are $(0.55,0.10,0.10,0.25)$. Each of the four category counts is generated by conditional Poisson splitting, either with $5H$ or with $5H$ times an independent gamma multiplier of shape two and scale one-half. Each reference is generated as in the third suite (untracked mixture for the intensity reference). At 2,000 and 10,000 cultures per arm, 100 repetitions were generated for each of two regimes, totaling 400 experiments. The influence functions for the intersection are combined on the same cultures, retaining covariance between the marginal and union estimates. At 10,000 cultures per arm, the matched designs recovered the $+15$ percentage points co-occurrence change with root mean squared error 0.71 and 0.73 percentage points for intensity and founder references. In the untracked mixture the intensity design recovered only $+7.16$ percentage points (root mean squared error 7.86 percentage points; zero coverage); the actual-founder design recovered $+15.06$ percentage points (root mean squared error 0.78 percentage points; 96\% coverage). Figure~\ref{fig:joint}a--c shows the exact pattern probabilities and recovered contrasts. The result demonstrates a change in joint outcomes that marginal frequencies cannot reveal; it is not a new biological finding.

All four suites together comprise 5,800 synthetic two-arm experiments. They assess finite-sample behavior, not uniform coverage, and use fixed mean calibrations. Independent seeding and reference misalignment are distinct from inter-founder competition. The identities allow arbitrary invariant within-founder mark distributions; heavy-tailed finite-sample performance, density-dependent competition and batch-dependent sampling remain untested here. Published fluctuation-analysis and $N$-mixture methods address related but not identical estimands; no claim of superiority to those methods is supported by these comparisons. Coverage differences of one or two percentage points with 100--200 repetitions are within ordinary Monte Carlo uncertainty.

\subsection{How accurate must calibration be?}
Even when marks are invariant, an imperfectly matched reference can change the recovered probability. Appendix~\ref{app:calibration} gives conservative outer bounds in terms of independently justified limits on reference mismatch and founder calibration. The bounds distinguish uncertainty about the observation model from ordinary sampling uncertainty. For the nominal 36\% and 32\% example, allowing independent five-percent errors in both calibrations in both arms gives a contrast range of $-10.84$ to $+2.80$ percentage points, even before sampling uncertainty. The nominal negative direction is then unresolved. For positive nominal estimates $0.36$ and $0.32$, an effective multiplicative calibration error preserves the negative sign if $k_e/k_0<1.125$ for the intensity design, or if $k_e/k_0<\log(0.64)/\log(0.68)\simeq1.157$ for the actual-founder design, where the multiplier acts on the integral $J(0)$. Under independent equal allowances on each of two calibration factors, these imply strict thresholds of 2.94\% and 3.65\%, respectively, before sampling uncertainty. A common positive scalar preserves direction in either design. Common reagents or even identical mismatch functions do not guarantee common effective scalars because different target laws induce different weights. These bounds do not correct exposure-dependent biology or interference between founders; a different experimental design is needed in those cases.

\section{A challenge can reveal a hidden repertoire}
Imagine two architectures that are indistinguishable under the entire baseline observation law, not merely the same measured mean. In our exact four-state example, two distinct latent states emit the same symbol. A continuous family of valid transition matrices preserves every baseline observation sequence, while the probability of entering one latent state ranges from zero to one-half. A calibrated challenge that responds only to that state yields exactly this hidden probability. Perfect specificity is not required: known response probabilities $a$ and $b$ in the two latent classes give $a+(b-a)p$, identifying $p$ if $a\ne b$. Uncalibrated responses may still be measured but do not identify the latent state probability. Figure~\ref{fig:joint}d illustrates the first transition. The construction in Appendix~\ref{app:challenge} makes the equality of baseline laws explicit. If the challenge were transformed along with the latent coordinates, it would merely change the representation. The independent physical calibration is essential. This illustrates why an informative intervention can reveal more than a longer series of the same measurements. It does not establish a general method for selecting challenges in arbitrary biological systems.

\section{How to use Pachinko in an experiment}
The accompanying software estimates founder-level outcome probabilities from aggregate counts, including joint patterns of two target outcomes. The larger framework also provides a way to plan lineage-resolved experiments and to stop when the required information is unavailable. Table~\ref{tab:workflow} specifies the workflow before fitting any model.
\begin{table}[!htb]\centering\small
\caption{A practical decision procedure. The software implements calibrated aggregate estimators, known-founder inference and necessary diagnostics, not a general repertoire prediction engine.}\label{tab:workflow}
\begin{tabularx}{\linewidth}{@{}p{1.25in}L@{}}\toprule
Step & Action and decision\\\midrule
Specify the target & Define the alteration, eligible founders, comparator, challenge, horizon and outcome. Distinguish formation, retention and detection.\\
Choose the measurement & If founders and their outcomes are directly tracked, estimate the founder-level contrast directly. If original founders can be counted but not linked, use the zero-outcome likelihood. Otherwise determine which calibrated reference architecture and mean founder count are feasible.\\
Check assumptions & Verify independent experimental units and randomized assignment where causal claims are intended. Check the within-arm conditional mean calibration, density-dependent response and competition. Separate scalar arm differences from differential mismatch.\\
Estimate the contrast & Supply one row per independent culture, arm, target count and reference count, and the calibrated mean founder number for each arm. Report the probabilities, difference, statistical uncertainty and calibration-sensitivity interval.\\
Resolve ambiguity & If calibration is missing, report bounds if justified, or a detected-founder conditional target only when a primary-founder count is available. Do not report a point estimate of unconditional capability from an invalid reference. If sensitivity is insufficient, distinguish uncertainty from absence. Select a challenge or assay aimed at the specific unresolved effect.\\
Test externally & Compare against direct lineage measurements and simpler matched-information estimators. Hold out entire perturbations or backgrounds to test generalization; random splitting of descendants is insufficient.\\\bottomrule
\end{tabularx}\end{table}

\subsection*{Choose the simplest measurement that answers the question}
If each founder can be individually linked to its outcomes, the direct fraction of positive founders is the natural estimate. When founders can be counted but their descendants cannot be linked, the zero-target culture still contains useful information: with $F$ original independent founders and a common positive-outcome probability $p$, the probability that none succeeds is $(1-p)^F$. Fitting this conditional probability across cultures estimates $p$ without a reference. This is the familiar limiting-dilution likelihood, including exact-count designs \cite{elda}; it does not require Poisson seeding. It does require that founder response is invariant across the measured counts and that founders act independently. If the original founder count cannot be observed at culture level, the appropriate intensity or actual-founder reference estimator is an alternative when its assumptions can be independently justified. If none of these measurement designs is defensible, a partial-identification interval or an explicit measurement gap is the appropriate output. These options are not competing algorithms with identical information: they trade additional physical measurements for weaker reliance on an aggregate observation model.

A practical pilot should include cultures with known original founder counts, a dilution series spanning the intended exposure range, independently measured reference response in each arm, and a test of whether target response per founder changes with density or reference introduction. A suitable assay must also count zero-outcome units, distinguish technical loss from biological extinction where relevant, and calibrate target detection for the endpoint. Nominal replication of thousands of cultures does not compensate for invalid calibration. An independent randomized culture is an experimental unit; shared donor or batch structure must be accounted for; descendants from one founder or multiple observations of the same culture are not new independent units.

\subsection*{Design for identification and precision}
Use a common founder stock, randomized plate positions, paired biological blocks and an independent calibration sample. Constant proportionality within each arm is required; the scale $c$ may differ between arms because normalization is arm-specific. Test target response across prespecified dilution levels with known seeding (or measured founder counts), with an explicit equivalence margin. A nonsignificant difference is not evidence of invariance. With two reference channels that track the same seeding quantity, normalized weighted reference moments $\E[Z_i s^Y]/\E[Z_i]$ should agree for all $s$; discrepancies can reject the joint assumptions, but agreement cannot establish them. The supplied diagnostics implement a prespecified grid and multiplicity-adjusted uncertainty. Under a conditional Poisson intensity reference only, the squared coefficient of variation of exposure is $\operatorname{CV}^2(H)=(\operatorname{Var}(Z)-\E[Z])/\E[Z]^2$. Extra reference dispersion invalidates this interpretation. When exposure variation is small, simpler estimators may be more efficient; there is no universal coefficient-of-variation cutoff.

Even under an ideal known-mean Poisson design with only positive or negative culture outcomes, information per culture about $p$ is $\lambda^2/(\exp(\lambda p)-1)$, maximized at $\lambda p\simeq1.594$. For $p_0=0.36$, $p_e=0.32$, a normal-approximation calculation with 80\% power and two-sided 5\% significance requires about 1,767 wells per arm at mean five founders. This is not a universal information bound for full count observations. The resource criterion matters: information per expected founder decreases with density. Our synthetic dilution pilot (eight known means from 0.5 to 8, 48 wells each, 1,000 repetitions per setting) had 4.7\% null rejection and only 7.4\% power for a log-linear decrease from 36\% to 32\%; even a decrease to 20\% was detected in 52.1\%. It is a feasibility design, not proof of equivalence. Figure~\ref{fig:design} summarizes the information and calibration trade-offs. Details and likelihoods appear in Appendix~\ref{app:inference}.
\begin{figure}[!htbp]\centering
\includegraphics[width=\linewidth]{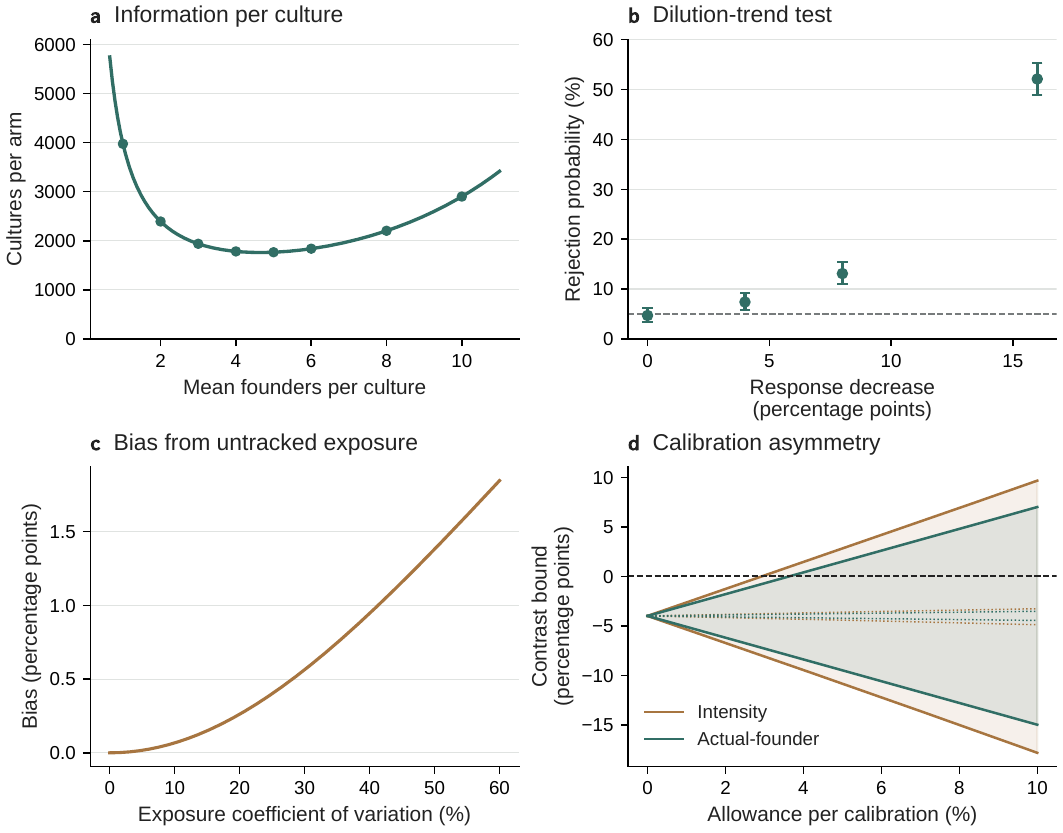}
\caption{Experimental design, information cost and calibration sensitivity. \textbf{a,} Normal-approximation number of independent cultures per arm for 80\% power at two-sided 5\% significance to detect a change from founder success probability 0.36 to 0.32. The design assumes Poisson seeding with exact known mean founder number and observes only whether any target is present. The minimum is near five founders per culture; the curve is not an efficiency bound for full-count estimators. \textbf{b,} Simulated dilution pilot: eight known Poisson means spanning 0.5 to 8 founders, 48 cultures at each level, 1,000 repetitions per response scenario. Points give the fraction of likelihood-ratio tests rejecting constant founder response; exact 95\% binomial intervals quantify Monte Carlo uncertainty, and the dashed line marks the nominal 5\% level. Response decreases are measured from 0.36 at the low end; the alternative is log-linear in log dilution. \textbf{c,} Exact population contrast bias of the homogeneous zero-count estimator for the symmetric exposure distribution $H=1\pm v$ with equal probabilities, mean five founders and coefficient of variation $v$. \textbf{d,} Conservative population bounds for the nominal probabilities 0.36 and 0.32. Shaded regions and solid boundaries allow independent errors in both arms, with the indicated allowance applied to each normalized reference-mean and founder-mean calibration factor. Ochre denotes the linear intensity identity, green the logarithmic actual-founder identity. Dotted curves show endpoint values under a shared positive scalar error. The horizontal dashed line marks zero contrast. These are sensitivity calculations, not sampling confidence intervals or measured calibration errors. The pilot is simulated, not an executed biological experiment.}\label{fig:design}
\end{figure}

\subsection*{A runnable example}
The reproducibility package contains a synthetic comma-separated values file with 10,000 independent cultures per arm and a minimal command-line estimator. The file has four columns: \texttt{arm}, \texttt{unit\_id}, \texttt{target\_count}, \texttt{reference\_count}. With Python and NumPy installed, from the package directory run:
\begin{verbatim}
python repro/pachinko.py repro/demo.csv \
    --mean-sham 5 --mean-operation 5
\end{verbatim}
This particular independent synthetic realization yields 35.37\% for sham, 31.90\% for operation, and a difference of $-3.47$ percentage points (nominal 95\% interval $-4.80$ to $-2.14$). The true probabilities are 36\% and 32\%. The script checks input units, reference-weighted zero counts, numerical integration stability and out-of-range estimates, and reports unresolved direction when its nominal interval overlaps zero or the probability estimates are invalid. It cannot verify biological assumptions from the data file. The interval assumes independent units and arms and known calibration; paired blocks and uncertain calibration require a different uncertainty calculation. The large number of cultures illustrates an information requirement, not a recommendation for ordinary laboratory scale.

Use the additional \texttt{--architecture founder} option only for a validated actual-founder reference. For two target channels the accompanying \texttt{joint\_estimator.py} uses the same culture-level influence functions to report the four binary founder patterns and the arm contrasts. To propagate example calibration bounds, save the estimator output as a JavaScript Object Notation (JSON) file and run \texttt{repro/sensitivity.py} with the four limits for each arm as described in the quick-start guide. In this example, hypothetical independent five-percent errors give a population plug-in outer interval of about $-10.2$ to $+3.3$ percentage points; including approximate simultaneous statistical uncertainty gives about $-12.4$ to $+5.4$ percentage points. The data do not determine these calibration limits: they must come from independent measurements or an explicitly declared sensitivity scenario. The tool also supplies a separate known-founder likelihood for experiments with the additional founder-count column.

\subsection*{Scope of applications}
Disease-specific hypotheses, biological instantiations and empirical analyses are reserved for separate manuscripts. The measurement definitions and mathematical results here do not rely on a particular genomic lesion, cell type or disease.

\section{What Pachinko establishes, and what comes next}
The contribution is a synthesis of an explicit biological target, two calibrated inverse calculations and their joint-outcome extension, exact examples of why familiar observations can mislead, and a procedure for choosing additional evidence. It is not a claim to have invented evolutionary contingency, marked Poisson processes, hidden-state models or partial identification. In particular, the restricted independent-secondary model is a mathematical reference case, not a proven reduction of recursive evolution. Its exact observation-channel identifiability and parameter fibers are supplied in Appendix~\ref{app:modelb} for readers who need to analyze that design.

The distinction between possible, realized and observed configurations survives even if the current estimator proves impractical. But that alone is not enough for a computational discovery method. The next test, outside this foundational article, is whether a real system admits either reference design, whether the method is efficient relative to established approaches adapted to the same estimand and information, and whether an independently measured challenge reveals a new dependency. The present study supplies a finite mathematical foundation, not evidence that those future tests will succeed. The practical conclusion is that an investigator should specify the original population, the outcome and the measurement design before interpreting a change in descendant abundance as a change in evolutionary capability.

\begin{table}[!htb]\centering\small
\caption{Relationship to established methods. The comparisons concern estimands and measurement assumptions, not untested performance rankings.}
\begin{tabularx}{\linewidth}{@{}p{1.48in}LL@{}}\toprule
Prior method & Established contribution & Relationship to this foundation\\\midrule
Fluctuation analysis \cite{luria,rosche} & Mutation events versus descendant counts & Distinct founder mark distribution and reference calibration.\\
Limiting dilution / ELDA \cite{elda} & Positive and negative cultures with known seeding & Prefer the known-founder likelihood when available.\\
Decompounding \cite{buchmann,lastpenrose} & Recovery of summand distributions & Classical generating function and size-bias ingredients; two reference architectures.\\
$N$-mixture models \cite{dennis,barker} & Latent abundance from replicated counts & Related fragility to heterogeneity and conditional observation assumptions.\\
Principal stratification and selection bounds \cite{frangakis,zhang,lee} & Post-treatment selection and truncated outcomes & Composite founder endpoints defined as zero after failure; not survivor-average effects.\\
Fixed-marginal causal bounds \cite{lu} & Partial identification of joint potential outcomes & Specific viable-switching objective.\\
Hidden-state identifiability \cite{blackwell,gilbert,jaeger} & Equivalent passive laws & Calibrated challenge example, not a new hidden-state theory.\\\bottomrule
\end{tabularx}\end{table}

\subsection*{Data and code availability} All new data are synthetic. The accompanying code includes the simulation studies, exact verification routines, figure-generation scripts and runnable aggregate estimators. No patient or unpublished institutional measurements are included.
\subsection*{AI assistance} Generative artificial intelligence tools (OpenAI ChatGPT, Google Gemini and Anthropic Claude) assisted with mathematical exploration, drafting, code development, numerical reanalysis and adversarial review. These tools are not authors. Responsibility for the scientific content, references, code and final submitted version rests with the author.

\clearpage\appendix\numberwithin{equation}{section}
\section{Founder outcome recovery: formal assumptions and proofs}\label{app:recovery}
The measurement problem is most transparent in a population of independent replicate cultures or experimental units. Let $H\geq0$ denote a latent exposure with $0<\E[H]<\infty$. Given $H$, the number of assigned founders is $F\sim\mathrm{Poisson}(\lambda H)$, where $\lambda>0$. Each founder produces a mark $(U,V)$ with $U\in\{0,1\}$ a primary detection indicator and $V\in\mathbb{N}_0$ the count of a specified secondary outcome. Marks are independent and identically distributed conditional on $F,H$, with a distribution independent of $H$. The entire within-founder branching history, including correlated survival and secondary production, may determine the mark. A failed founder may contribute $(0,0)$. The aggregate measurements are $X=\sum_{i=1}^F U_i$ and $Y=\sum_{i=1}^F V_i$.

A nonnegative reference measurement $Z$ satisfies $\E[Z\mid H,Y]=cH$, $c>0$. Conditional independence of $Z$ from the target clade given $H$, together with $\E[Z\mid H]=cH$, suffices. The conditional mean restriction alone is enough for the two generating function identities; a conditional Poisson distribution for $Z$ is not required. These assumptions concern the experimental unit: a reference affected by treatment, biological feedback or distinct exposure cannot be presumed valid. All results concern the specified observed mark, which may differ from a latent biological outcome.

\begin{proposition}[Reference-normalized detected-founder distribution]
If $\E[U]>0$, then for $0\leq y\leq1$, with $0^0=1$,
\begin{equation}\label{eq:conditional}
 \E[y^V\mid U=1]=\frac{\E[X y^Y]}{\E[Z y^Y]}\frac{\E[Z]}{\E[X]}.
\end{equation}
The identity permits arbitrary dependence of $U$ and $V$ within a founder and arbitrary shared exposure distribution satisfying the stated mean condition.
\end{proposition}
\begin{proof}
Write $\phi(y)=\E[y^V]$. By conditioning on the Poisson founder count, the joint generating function of $X,Y$ given $H$ is $\exp\{\lambda H(\E[x^U y^V]-1)\}$. Differentiation at $x=1$ gives $\E[Xy^Y\mid H]=\lambda H\E[Uy^V]\exp\{\lambda H(\phi(y)-1)\}$. The conditional mean restriction yields $\E[Zy^Y\mid H]=cH\exp\{\lambda H(\phi(y)-1)\}$. After averaging, their ratio is $\lambda\E[Uy^V]/c$. The factor $\E[Z]/\E[X]=c/(\lambda\E[U])$ completes the proof. The common exposure factor is strictly positive and finite for all $y\in[0,1]$. Equality of generating functions determines the count law. For vector-valued marks substitute $\prod_k y_k^{V_k}$ for $y^V$ throughout.
\end{proof}
At $y=0$, equation~\eqref{eq:conditional} identifies the probability of no observed secondary outcome among detected primary founders; its complement identifies the probability of at least one. The derivative at one, when the necessary moments exist, gives the conditional expected yield. In particular, if second moments are finite, it equals
\begin{equation}
 \E[V\mid U=1]=\frac{\operatorname{Cov}(X,Y)}{\E[X]}-\frac{\operatorname{Cov}(Y,Z)}{\E[Z]}.
\end{equation}
This follows by differentiating the ratio in equation~\eqref{eq:conditional} at one; the two products of marginal means cancel. The subtraction removes the shared-exposure contribution, not within-founder survival dependence. Under the stronger condition $\E[Z\mid H,X,Y]=cH$, total covariance also implies $\operatorname{Cov}(X,Z)/\E[X]=\operatorname{Cov}(Y,Z)/\E[Y]$ when $\E[Y]>0$. Thus the alternative subtraction $(\E[Y]/\E[Z])\operatorname{Cov}(X,Z)/\E[X]$ is valid under that stronger condition. It is not valid merely from $\E[Z\mid H,Y]=cH$.

For an explicit counterexample to the shortcut, let $H=1$, $F\sim\operatorname{Poisson}(1)$, $V_i=1$ and $U_i\sim\operatorname{Bernoulli}(1/2)$. Then $Y=F$, $X\mid Y\sim\operatorname{Binomial}(Y,1/2)$ and $Z=1+\mathbf{1}\{X>0\}-(1-2^{-Y})$ is nonnegative with $\E[Z\mid H,Y]=1$. Here $\operatorname{Cov}(Y,Z)=0$ but $\operatorname{Cov}(X,Z)=\E[X2^{-Y}]=e^{-1/2}/4>0$. The ratio derivative correctly gives one; replacing the $Y,Z$ covariance by the scaled $X,Z$ covariance incorrectly reduces it. Conditional yield is not a causal tolerance probability.

\begin{proposition}[Calibrated unconditional founder distribution]\label{prop:uncond}
If $m=\lambda\E[H]>0$ is independently calibrated, the unconditional generating function is identified by
\begin{equation}\label{eq:integral}
 \phi(y)=1-\frac{\E[Z]}{m}\int_y^1\frac{\E[Ys^{Y-1}]}{\E[Zs^Y]}\,\mathrm{d}s,\quad 0\leq y\leq1.
\end{equation}
Terms with $Y=0$ in the numerator are defined as zero. The integral at endpoints is understood by a limit.
\end{proposition}
\begin{proof}
For $0<s<1$, differentiate $\E[s^Y\mid H]=\exp\{\lambda H(\phi(s)-1)\}$. The numerator in (\ref{eq:integral}) is
\[\lambda\phi'(s)\E[H\exp\{\lambda H(\phi(s)-1)\}],\]
and the denominator is $c$ times the same expectation. Hence their ratio is $\lambda\phi'(s)/c$. Multiplying by $\E[Z]/m=c/\lambda$ and integrating gives $1-\phi(y)$. Nonnegative terms justify the endpoint limit by monotone convergence of the series defining $\phi$. No finite first moment of $V$ is needed for the generating function identity itself; statistical inference below uses finite moments.
\end{proof}

\begin{proposition}[Actual-founder reference for arbitrary seeding]\label{prop:founder}
Let $F$ have any nonnegative integer distribution with $0<m=\E[F]<\infty$, and let the marks $(U_i,V_i)$ conditional on $F$ be independent and identically distributed with a law independent of $F$. If $\E[Z\mid F,Y]=cF$ with $c>0$, then for $0<y\leq1$,
\begin{equation}\label{eq:logrecovery}
 \phi(y)=\exp\left\{-\frac{\E[Z]}{m}\int_y^1\frac{\E[Ys^{Y-1}]}{\E[Zs^Y]}\,\mathrm{d}s\right\}.
\end{equation}
The value at zero is the right limit, with $\exp(-\infty)=0$. If $\E[U]>0$, the detected-founder law satisfies
\begin{equation}\label{eq:founderconditional}
 \E[y^V\mid U=1]=\phi(y)\frac{\E[Xy^Y]}{\E[Zy^Y]}\frac{\E[Z]}{\E[X]}.
\end{equation}
\end{proposition}
\begin{proof}
Let $G_F(x)=\E[x^F]$. Conditioning on $F$ gives $G_Y(s)=G_F(\phi(s))$. For $0<s<1$,
\begin{align*}
 \E[Ys^{Y-1}]&=G'_F(\phi(s))\phi'(s),\\
 \E[Zs^Y]&=c\phi(s)G'_F(\phi(s)),\\
 \E[Xs^Y]&=G'_F(\phi(s))\E[Us^V].
\end{align*}
Here $\phi(s)>0$ and $G'_F(\phi(s))>0$ because $m>0$. Thus the first two expressions give $\phi'(s)/(c\phi(s))$, which integrates to $-\log\phi(y)/c$ using $\phi(1)=1$. Since $\E[Z]=cm$, equation~\eqref{eq:logrecovery} follows. The third expression, normalized by $\E[X]=m\E[U]$, proves equation~\eqref{eq:founderconditional}. Endpoint values follow by continuity of the generating functions. With finite required moments, differentiating equation~\eqref{eq:founderconditional} gives $\E[V\mid U=1]=\E[V]+\operatorname{Cov}(X,Y)/\E[X]-\operatorname{Cov}(Y,Z)/\E[Z]$. At zero a denominator may vanish; the displayed ratio then has to be interpreted by its limit, not by dividing zeros. If $\phi(0)=0$, the integral diverges and the positive-outcome probability is one. No finite mean of $V$ is required for the population identity; statistical regularity is separate.
\end{proof}
A reference proportional to $F$ is not an intensity reference: under Poisson seeding it also covaries with the realized number of founders at fixed $H$. Conversely, a valid intensity reference does not generally satisfy the actual-founder conditional mean. If mean founder number is unknown, the grouping of original founders is itself ambiguous: one model can have $F$ founders with marks equal to the sum of $k$ independent and identically distributed elementary marks and another $kF$ founders with the elementary mark law, for integer $k>1$. Both have identical $Y$ and can have the same conditional reference law by scaling $c$ by $1/k$ in the second model. Their per-founder zero probabilities generally differ. This is a nonidentifiability witness for the aggregate $Y,Z$ design.

\subsection{Joint recovery and binary outcome patterns}
Assume the reference conditional mean restriction holds with the full vector $\bm Y$ in place of $Y$. Let $\bm V\in\mathbb N_0^d$ and $\bm Y=\sum_i\bm V_i$; define $\phi(\bm y)=\E[\prod_{j=1}^d y_j^{V_j}]$. For $\bm y\in(0,1]^d$, set $\bm x(t)=\bm 1+t(\bm y-\bm 1)$, $0\le t\le1$, and
\[
 D(t)=\frac{\mathrm{d}}{\mathrm{d}t}\E\left[\prod_j x_j(t)^{Y_j}\right],\qquad
 B(t)=\E\left[Z\prod_j x_j(t)^{Y_j}\right].
\]
Differentiation along the path gives $D(t)/B(t)=(\lambda/c)\,\mathrm{d}\phi(\bm x(t))/\mathrm{d}t$ in the intensity design and $D(t)/B(t)=(1/c)\,\mathrm{d}\log\phi(\bm x(t))/\mathrm{d}t$ in the actual-founder design. Integration from zero to one yields, respectively,
\[
 \phi(\bm y)=1+\frac{\E[Z]}{m}\int_0^1\frac{D(t)}{B(t)}\,\mathrm{d}t,
 \qquad
 \log\phi(\bm y)=\frac{\E[Z]}{m}\int_0^1\frac{D(t)}{B(t)}\,\mathrm{d}t.
\]
The path is nonincreasing coordinatewise, permitting endpoint limits; derivatives are on the interior where needed. Coordinates that remain one contribute zero to the path derivative. The same proof applies to any subset sum $Y_S=\sum_{j\in S}Y_j$ because the full-vector reference condition implies the scalar condition by iterated expectation. For $S\subseteq\{1,\ldots,d\}$ write $h(S)=\Pp(V_j=0\ \forall j\in S)$, with $h(\varnothing)=1$. The probability of exactly the positive set $A$ is
\begin{equation}
 \Pp(\{j:V_j>0\}=A)=\sum_{B\subseteq A}(-1)^{|B|}h(A^c\cup B).
\end{equation}
This follows by expanding the product of indicators of positive events on $A$. The finite-sample estimates need not satisfy all simplex constraints; report incompatibilities and do not silently clip. Estimation of all $2^d$ patterns is not scalable by this formula alone. The executed numerical benchmark validates two binary channels, not arbitrary high-dimensional generating function reconstruction.

This result makes failed establishment part of the estimand when calibration counts all originally assigned founders. For matched experiments with operation and sham, the probability contrast of at least one counted target event is
\begin{equation}\label{eq:contrast}
 \Delta_e C_T=\phi_0(0)-\phi_e(0).
\end{equation}
With randomized assignment and appropriate treatment consistency, absence of interference, common target definition and calibrated observation, this is a founder-level causal contrast for the declared event. Without those conditions it is a contrast between observed mark laws. Even randomized treatment does not turn a detection endpoint into an ever-attainment endpoint. Separate calibration is needed if these differ.

\subsection{Calibration is necessary, not decorative}
The mean-founder calibration addresses a genuine nonidentifiability. Replace $\lambda$ by $a\lambda$ for any $a>1$ and the founder mark law by its mixture with an all-zero mark, retaining the original mark with probability $1/a$. The conditional aggregate generating function of $X,Y$ is unchanged at every $H$, and if the reference is generated independently of target marks given $H$ by the same conditional law in both worlds, the complete law with reference $Z$ is unchanged. This submodel suffices for the nonidentifiability witness. Yet the unconditional positive-outcome probability decreases by $1/a$. No amount of observation of those aggregate channels alone can distinguish these worlds. In contrast, equation (\ref{eq:conditional}) is invariant to this transformation: the conditional detected-founder target does not require the calibration. This illustrates target-specific identification.

If $m$ is known only to lie in $[m_L,m_U]$, the positive-outcome probability lies in $[A/m_U,A/m_L]\cap[0,1]$, where $A$ denotes the measured integral in (\ref{eq:integral}) multiplied by $\E[Z]$. These are necessary bounds; sharpness requires that the proposed calibrations are compatible with all available marks and constraints. For two arms, marginal intervals imply an outer bound on the contrast by subtracting opposite endpoints. It is sharp only when the two extrema are jointly attainable. If a counted positive outcome has no false positives and conditional sensitivity is bounded below by $d_L>0$, the latent positive-outcome probability lies between the counted probability and its value divided by $d_L$ (capped at one). This second bound is a distinct measurement problem.

\subsection{A stronger reference assumption gives a simpler estimator}
When $Z\mid H\sim\mathrm{Poisson}(cH)$, write $G_Z(t)=\E[t^Z]$ and $G_Y(y)=\E[y^Y]$. Then
\begin{equation}\label{eq:match}
 G_Y(y)=G_Z\left(1-\frac{\lambda}{c}(1-\phi(y))\right).
\end{equation}
If $c\geq\lambda$, the argument lies in $[0,1]$. Strict monotonicity of $G_Z$ on the interior gives a direct numerical inversion. This is a natural matched-information comparator to (\ref{eq:integral}), using the same counts and calibration but requiring the full conditional Poisson reference law. Neither method claims to be an optimal estimator. For homogeneous known exposure, standard decompounding gives $1-\phi(0)=-\log\Pp(Y=0)/m$. The reference-based identities address common unknown exposure without assuming its distribution is known.

\section{Calibration sensitivity and alternative measurements}\label{app:calibration}
Let $W$ be an arbitrary nuisance vector with effective exposure $h(W)\geq0$, normalized so that $\E[h(W)]=1$. Conditional on $W$, let $F\sim\operatorname{Poisson}(m h(W))$ and let founder marks have a common distribution independent of $W$. Let $M_Z=\E[Z]>0$. Suppose the conditional mean of the reference is
\begin{equation}
 \E[Z\mid W,Y]=M_Z h(W) r(W),\qquad \E[h(W)r(W)]=1,
\end{equation}
where $0<L\leq r(W)\leq U<\infty$ on positive exposure. This describes imperfect proportionality in effective exposure, not arbitrary dependence of the reference on the target outcome. The true mean founder number $m$ may differ from the entered value $m_{\mathrm{cal}}$. Write $l\leq m/m_{\mathrm{cal}}\leq u$ with $0<l\leq u$. These bounds are externally supplied.

\begin{proposition}[Conservative calibration bounds]
Let $A_{\mathrm{cal}}$ denote the population value of the positive-outcome integral with entered calibration $m_{\mathrm{cal}}$. The true positive-outcome probability $p=1-\phi(0)$ obeys
\begin{equation}\label{eq:mismatch}
 p\in\left[\frac{L A_{\mathrm{cal}}}{u},\frac{U A_{\mathrm{cal}}}{l}\right]\cap[0,1].
\end{equation}
\end{proposition}
\begin{proof}
For $s\in(0,1)$, let $G_W(s)=\exp\{mh(W)(\phi(s)-1)\}$. The derivative numerator is $m\phi'(s)\E[h(W)G_W(s)]$, and the reference denominator is $M_Z\E[h(W)r(W)G_W(s)]$. Therefore
\begin{align*}
 \overline r(s)&=\frac{\E[h(W)r(W)G_W(s)]}{\E[h(W)G_W(s)]}\in[L,U],\\
 A_{\mathrm{cal}}&=\frac{m}{m_{\mathrm{cal}}}\int_0^1\frac{\phi'(s)}{\overline r(s)}\,\mathrm{d}s.
\end{align*}
Since $\phi'(s)\geq0$ and $\int_0^1\phi'(s)\,\mathrm{d}s=p$, we obtain $l p/U\leq A_{\mathrm{cal}}\leq u p/L$. Rearranging gives the result. Endpoints are defined by nonnegative integration as in Proposition~\ref{prop:uncond}.
\end{proof}
These outer bounds need not be sharp: the extremizing reference functions must also satisfy normalization and the complete observed law. If the reference and founder errors are dependent across arms, imposing that joint structure can tighten the contrast interval. Given arm intervals $[p_{0,L},p_{0,U}]$ and $[p_{e,L},p_{e,U}]$, a conservative contrast interval is $[p_{e,L}-p_{0,U},p_{e,U}-p_{0,L}]$. Statistical uncertainty can be propagated by constructing simultaneous intervals for the two nominal integrals and applying the same monotone transformations. The supplied tool uses Bonferroni-adjusted, approximate Wald intervals, not exact finite-sample confidence regions. An empty intersection with $[0,1]$ signals incompatibility of the estimate or calibration bounds; it is not a valid probability estimate.

\subsection{Sensitivity for the actual-founder reference}
For the actual-founder design, suppose $\E[Z\mid F,Y]=M_Z(F/m)r(F)$ with $L\le r(F)\le U$ for positive $F$ and $\E[Fr(F)]=m$. Let $l\le m/m_{\mathrm{cal}}\le u$. With $J_{\mathrm{cal}}$ the population value of the integral multiplied by $M_Z/m_{\mathrm{cal}}$, the same weighted-mean argument gives
\[
 \frac{l}{U}[-\log\phi(0)]\le J_{\mathrm{cal}}\le
 \frac{u}{L}[-\log\phi(0)].
\]
Indeed, the denominator is $M_Z\E[Fr(F)\phi(s)^F]/m$ and the numerator is $G'_F(\phi(s))\phi'(s)$, so the normalized reference distortion is a weighted mean of $r(F)$. Exponentiating gives
\begin{equation}\label{eq:logbounds}
 1-\exp(-LJ_{\mathrm{cal}}/u)\le p\le1-\exp(-UJ_{\mathrm{cal}}/l).
\end{equation}
The endpoint $p=1$ follows by a limit. As with the intensity case these are outer bounds, not generally sharp identified sets. They address only bounded proportionality error, not mark dependence on $F$. The same argument works with additional nuisance variables if marks remain invariant and the conditional mean is bounded relative to actual founder number.

\subsection{Known-founder likelihood}
Following limiting-dilution analysis \cite{elda}, if the original number of founders $F_i$ in culture $i$ is observed, define $D_i=\mathbf{1}\{Y_i>0\}$. Independent invariant founder marks imply $\Pp(D_i=0\mid F_i)=(1-p)^{F_i}$ without any assumption on the distribution of $F_i$. For independent cultures the log likelihood is
\begin{equation}
 \ell(p)=\sum_i\left[D_i\log\{1-(1-p)^{F_i}\}+(1-D_i)F_i\log(1-p)\right].
\end{equation}
Cultures with $F_i=0$ provide no information, and a positive target in such a culture is inconsistent with the specified model. Interior estimation can use the scalar score equation. The conditional expected information is
\begin{equation}
 \mathcal{I}(p)=\sum_{i:F_i>0}\frac{F_i^2(1-p)^{F_i-2}}{1-(1-p)^{F_i}}.
\end{equation}
This is obtained from the Bernoulli success probability and its derivative. Boundary estimates require non-Wald inference. For fully observed founder outcomes the natural estimate is the total number of positive founders divided by the total number of original founders; culture-level uncertainty accounts for the experimental unit. None of these estimators recovers an unobserved biological endpoint without an observation model.

\section{Branching, proposal and acceptance: formal examples}\label{app:branch}
The outcome law is more general than a particular tolerance model. To make the difference explicit, consider a restricted reference model with primary retained-founder counts $F_j\sim\mathrm{Poisson}(\Lambda g_jt_j)$; each founder independently emits a primary Bernoulli observation with probability $r_j=s_jd_j$ and Poisson secondary observations with mean $\eta_{jk}=g_kt_{k\mid j}s_kd_k$. Here $g,t,s,d$ are generation, tolerance, establishment and detection probabilities in $(0,1)$; $\Lambda$ is known. We denote this restricted compound-Poisson count model by Model B, to distinguish it from the more general recovery identities in Appendix~\ref{app:recovery}. The conditional factorization is deliberately restrictive. In the homogeneous model,
\begin{equation}
 \mu_j=\lambda_jr_j,\quad \nu_{jk}=\lambda_j\eta_{jk},\quad c_{jk}=\lambda_jr_j\eta_{jk},\quad \operatorname{Var}(O_{jk})-\nu_{jk}=\lambda_j\eta_{jk}^2.
\end{equation}
For $\mu_k>0$ and identifiable $\eta_{jk}$, the observable ratio $\Lambda\eta_{jk}/\mu_k$ equals $t_{k\mid j}/t_k$ under those exact shared-factor assumptions. The covariance form $\Lambda c_{jk}/(\mu_j\mu_k)$ additionally requires $\mu_j>0$. In particular the use of the same $g_k,s_k,d_k$ in primary and secondary contexts is essential. Appendix~\ref{app:modelb} records the joint law, observation-channel identification and joint tolerance bounds. This benchmark is not a generic consequence of biological parentage.

If rejection leaves the process unchanged, accepted operations of proposal intensity $\gamma$ and acceptance probability $\tau$ occur at intensity $\gamma\tau$. Reciprocal scaling $(\gamma,\tau)\mapsto(a\gamma,\tau/a)$ preserves the accepted transition generator for admissible, nonexplosive processes. This follows by thinning \cite{lastpenrose}. The result holds for branching histories as well as terminal counts. Consequential rejection can break the symmetry: for a single operation with lethal rejection and all other death rates fixed, changing the rejected intensity changes survival. It does not necessarily break it when changes in other transition fluxes compensate. Measuring total death without its causes need not identify the rejected-proposal rate.

For an exact branching counterexample, let each retained primary founder begin a critical linear birth--death process with birth and death rates one, at time zero. At time $u$ the population size is $L(u)$. Secondary events are nondepleting Poisson arrivals proportional to living-cell time and generate immortal detectable targets. Observation of a primary founder requires survival at $T=1$. By the branching property the probability of survival given $L(u)=n$ is $1-q(1-u)^n$, where $q(t)=t/(1+t)$. The critical branching generating function is $f_u(z)=1-(1-z)/(1+u(1-z))$. Differentiating gives
\begin{equation}
 \E[L(u)\mid L(1)>0]=1+u+\frac{u(1-u)}{2},\qquad
 \E\left[\int_0^1 L(u)\,\mathrm{d}u\mid L(1)>0\right]=\frac{19}{12}.
\end{equation}
The unconditional expected occupation time is one. With equal primary and secondary reference opportunity durations of one time unit, the same calibrated generation rate and detection, a true relative acceptance probability of $4/5$ becomes an apparent covariance ratio of $(4/5)(19/12)=19/15$. In this construction with survival-gated detection and immortal targets, the direction reverses even with zero net growth and homogeneous exposure. For a linear birth--death model with nondepleting secondary arrivals and an independent secondary survival kernel, the occupation factor is an integral of survivor-conditioned population size weighted by that kernel at the remaining clone age; its time units must be matched to the primary reference opportunity. It can equal one in special cases. Exposure normalization alone is insufficient under this observation model. Independent occupation calibration is needed.

An additional independent source of distortion occurs if founder exposure is variable. For founder marks with means $p=\E[U]$, $v=\E[V]$ and joint moment $w=\E[UV]$, $\E[H]=1$, and $\operatorname{Var}(H)=\sigma_H^2$, total covariance gives $\operatorname{Cov}(X,Y)=\lambda w+\sigma_H^2\lambda^2pv$. With the critical example above, unconditional secondary yield relative to the unit reference is $4/5$ and survivor-conditioned yield is $19/15$. Choosing $\lambda=5$ and $\sigma_H^2=1/4$ adds one to the ratio, giving $34/15$ before correction. Reference normalization removes the shared-exposure term and leaves $19/15$; division by the calibrated occupation factor recovers $4/5$. Figure~\ref{fig:performance}a displays these exact model quantities. Without the occupation calibration and shared factorization, the conditional outcome law remains interpretable but the tolerance does not.

A more direct contrast shows why tolerance is not the target capability. In sham, two viable daughters each have target-event probability $1/5$, giving positive outcome probability $9/25$. An operation doubling daughter-level probability to $2/5$ yields $16/25$ if establishment is certain, but $8/25$ if the operation establishes in only half of assigned founders. The corresponding changes are $+7/25$ and $-1/25$. An identical twofold conditional increase can therefore have opposite effects on the founder-level repertoire.

\subsection{General reversal conditions}
If a positive endpoint requires establishment, the product criterion in the main text is necessary and sufficient regardless of subsequent dependence. If a founder can generate a positive endpoint before failing establishment, an additional term $(1-a_e)q^{\mathrm{fail}}_e$ must be included; the product criterion alone no longer holds. Conditional on establishment, if $K_e$ daughters have independent identically distributed successes with probability $s_e$ conditional on $K_e$, then $q_e=1-G_{K_e}(1-s_e)$. This permits different daughter-number laws across arms. For shared founder propensity $S_e$, with conditional independence given $K_e,S_e$, use $q_e=\E[1-(1-S_e)^{K_e}\mid\text{establishment},e]$. For fixed $K_e=k\ge1$ and common marginal daughter probability $s_e$ but unrestricted dependence, $s_e\le q_e\le\min(1,ks_e)$ by event containment and the union bound. Both endpoints are attainable (identical events or events arranged to cover as much probability mass as possible). Thus marginal daughter probabilities alone cannot identify $q_e$ without dependence information. These are finite-horizon endpoint statements, not a general fitness ordering.

\subsection{Histories and their representations}
Let $\mathcal{M}$ be a finite set of admissible histories, $M$ a random history on that set under a specified prior, and $R=f(M)$ its deterministic representation. A representation $r$ defines a fiber $f^{-1}(r)$. Use natural logarithms (entropy is in nats). Under a uniform prior on $N$ histories, a fiber of size $n_r$ has probability $n_r/N$ and conditional entropy $\log n_r$. Under a nonuniform prior the count alone does not determine either quantity. For any prior, when $\Pr(R=r)>0$, Bayes' rule gives $D_{\rm KL}(P(M\mid R=r)\Vert P(M))=-\log\Pr(R=r)$. This follows because the posterior-to-prior ratio on the fiber is constant. With stochastic detection $Q(r\mid m)$, the posterior instead reweights histories by $Q$; a raw path count is not an inference of the realized history. This elementary distinction is the history--representation component of Pachinko, rather than a claim of new information theory.

\section{Restricted count-model identifiability and target-specific bounds}\label{app:modelb}
For the restricted model in the strict interior in Appendix~\ref{app:branch} (known $\Lambda>0$, $\lambda_j>0$, $0<r_j<1$, and $\eta_{jk}>0$ for observed pairs), let $P$ denote the set of observed ordered pairs, $K$ the number of classes, and $z_0$ the number of vertices without observed outgoing edges. For a source $j$, the secondary Poisson rates are independent of the primary Bernoulli mark conditional on one retained founder, and the block generating function is
\[G_j(x,\mathbf y)=\exp\{\lambda_j[(1+r_j(x-1))\exp(\sum_{k:(j,k)\in P}\eta_{jk}(y_k-1))-1]\}.\]
Blocks are independent by construction. First and second factorial cumulants give the equations in Appendix~\ref{app:branch}; if at least one outgoing edge is observed with positive intensity, $\eta=c/\mu$, $r=c/\nu$, $\lambda=\mu\nu/c$. Alternatively $\eta=(\operatorname{Var}(O_{jk})-\nu)/\nu$ and $\lambda=\nu/\eta$; $r$ follows from $\mu$. At a sink only $\mu=\lambda r$ is identified. These recovered quantities determine the displayed generating function, so equality of the invariants is both necessary and sufficient for equality of the complete observation laws, not merely of selected moments. Self-loops are allowed provided primary and secondary counts are distinct measured channels. Incoming edges constrain feasible parameters but do not identify $\lambda,r$ separately in the interior: for scale $b$ near one, multiply $t_k$ and all incoming conditional tolerances by $b$ and divide $s_k$ by $b$.

The observation log-map is affine, a special case of established parameter-redundancy analysis \cite{catchpole}. Each edge row has a unique conditional tolerance pivot. Each nonsink contributes two independent primary product rows and each sink one. Hence rank is $2K-z_0+|P|$, and the fiber dimension in the full $4K+K^2$ parameter space is $2K+z_0+K^2-|P|$. In original coordinates the equivalence set is the exponential of an affine slice in the negative orthant, not a linear vector space. Row-space membership is necessary and sufficient for point identification of a log-linear target on this interior domain: if not in the row space, a kernel displacement changes the target while remaining feasible for sufficiently small steps. At boundary points the interior rank statement does not apply. In the reduced model with $\lambda_j>0$, an outgoing channel with $\nu_{jk}>0$ identifies its rate and the founder intensity by second factorial cumulants, even if $r_j=0$ or $1$. All other declared outgoing rates then follow, including zero rates. If all secondary means vanish, they identify zero secondary rates within the positive-$\lambda_j$ class, but the primary law identifies only $\mu_j$; it does not in general separate $\lambda_j$ and $r_j$. If $\lambda_j$ is allowed to vanish and the entire block is zero, zero marks and absent founders cannot generally be distinguished. A positive primary mean alone also certifies positive $\lambda_j$, so zero observed secondary rates are then identified. Thus the ratio based on $\eta_{jk}$ requires $\mu_k>0$ and identifiable $\eta_{jk}$; covariance additionally requires $\mu_j>0$. A zero target mean does not establish zero biological tolerance. Missing channels are not zero-rate channels.

For a block with at least one secondary channel, full support on the entire nonnegative integer lattice holds exactly when $\lambda_j>0$, $0<r_j<1$ and every declared secondary rate is positive. Sufficiency follows by choosing enough founders to emit each desired count, using unobserved-primary founders for secondary events when needed. Necessity follows from the unavailable primary or secondary positive outcomes at zero rates, or from the impossibility of $O_j=0$ with positive secondary count when $r_j=1$. At a sink the condition is instead $\lambda_jr_j>0$, including $r_j=1$. On a fixed boundary stratum, zero rates and fixed primary probabilities do not add free coordinates; we do not claim a universal boundary dimension. Sample zeros are not structural zeros. At the upper biological corner $\mu_j=\Lambda$ forces all four primary probabilities to one, so the interior sink nonidentifiability statement cannot be extended unconditionally to the boundary. When an observed target has $r_k=0$, a zero incoming rate may hold for every conditional tolerance; counting each zero edge as an independent fiber constraint is therefore incorrect. For example with two classes and one declared edge $1\to2$, positive founder intensities and $r_2=0$, the law fixes the primary product in class 1 and forces $s_2d_2=0$ but does not constrain $t_{2\mid1}$. In the full twelve-coordinate closed cube the maximal fiber dimension is ten (three free primary factors in each block and four free edge coordinates), not nine. A universal boundary-dimension formula is not asserted.

For a target $k$, put $a=\mu_k/\Lambda$, $q=g_ks_kd_k$, $M=\max(a,\{\eta_{jk}:(j,k)\in P\})$. The joint observed tolerance vector equals $(a,\{\eta_{jk}\})/q$. Its sharp domain is $q\in(M,U)$ where $U=r_k$ if $k$ has an outgoing edge and $U=1$ otherwise. Necessity follows from probabilities strictly below one. 

For a nonsink fix $r_k$, take $g_k=q/r_k$, $t_k=a/q$, $t_{k\mid j}=\eta_{jk}/q$, and $s_k=d_k=\sqrt{r_k}$. This also gives the required $\lambda_k=\mu_k/r_k$. At a sink choose any $r_k\in(q,1)$ and apply the same construction. All unobserved edge parameters can be chosen independently in $(0,1)$. Thus every point of the ray segment is feasible, endpoints approached but not attained. Ratios are identified even though absolute scales are not. Measuring one appropriate common scale can resolve all incoming tolerances, without determining the entire fiber. In particular, a primary tolerance calibration fixes $q=a/t_k$. For a sink, the separate reduced product $r_k$ ranges over $(M,1)$ and $\lambda_k=\mu_k/r_k$ ranges over $(\mu_k,\mu_k/M)$. Incoming observed rates can therefore weakly tighten sink bounds even without restoring point identification. An added rate strictly tightens them precisely when it exceeds the previous maximum $M$; relative to no incoming edge this means $\eta_{jk}>a$, or $R_{jk}>1$. The same criterion tightens the absolute tolerance segment at a nonsink. 

More explicitly, the entire interior fiber is parameterized by $g_k,s_k$ for nonsinks, $r_k,g_k,s_k$ for sinks, and all unobserved conditional tolerances. At a nonsink choose $g_k\in(M/r_k,1)$ and $s_k\in(r_k,1)$; at a sink choose $r_k\in(M,1)$, then the same intervals. Reconstruct $t_k=a/(g_kr_k)$, $d_k=r_k/s_k$ and $t_{k\mid j}=\eta_{jk}/(g_kr_k)$ for observed pairs. This chart is global: projection onto the free coordinates is its inverse. Its log-coordinate inequalities are affine, establishing connectedness and the stated dimension on the open domain. 

The observable image is exactly the product of target blocks with $0<\lambda_k<\Lambda$, $0<r_k<1$ and $0<\eta_{jk}<r_k$ into nonsinks, and $0<\mu_k<\Lambda$, $0<\eta_{jk}<1$ into sinks. These conditions are sufficient by the preceding reconstruction; a violated inequality excludes this factorization. The dimension decomposes into $2K$ biological factorization directions, $K^2-|P|$ unobserved coordinates and $z_0$ sink collapse directions; the first two depend on parametrization and observation scope, not on a universal biological law. The exact projection of an unobserved $\eta_{jk}$ is $(0,r_k)$ into a nonsink or $(0,1)$ into a sink; the latter is coupled to $r_k$ and not a Cartesian product. For the observed absolute tolerances the sharp intervals are $(a/U,a/M)$ and $(\eta_{jk}/U,\eta_{jk}/M)$. At a nonsink, $g_k\in(M/r_k,1)$ and $s_k,d_k\in(r_k,1)$; at a sink each of $g_k,s_k,d_k$ projects to $(M,1)$. These follow by taking the other free factors arbitrarily close to one. The full ray segment, not independent combinations of its projections, is the identified set. All statements assume a product domain; imposing a shared proposal simplex changes feasibility and may couple targets.

For a target-specific assay, direct measurement of $t_k$ identifies every observed incoming $t_{k\mid j}=R_{jk}t_k$, even at a sink. At a nonsink, generation $g_k$ is an equivalent anchor because $\lambda_k$ is known; at a sink, $g_k$ alone does not resolve $g_kr_k$. The rank test for a log-linear target is applied after appending rows corresponding to independently calibrated stage products. In the interior, full separation of the four primary probabilities requires two independent scalar stage measurements at a nonsink (one of $g,t$ and one of $s,d$), or three independent single-factor measurements at a sink. This is a statement about the specified observation law, not a prescription to treat an experimental perturbation as a direct stage measurement. A calibrated composite can replace multiple single-factor observations if it supplies the appropriate row-space information.

The closure is obtained by replacing strict inequalities by non-strict ones, because an interior point and any closed-domain point can be interpolated in log coordinates for the observed positive products; unobserved probabilities have independent closed intervals. Positive observed primary products keep the reconstruction denominators bounded away from zero.

\subsection{Competing single-factor explanations}
The observable ratio $Q_{jk}=\Lambda\eta_{jk}/\mu_k$ need not measure tolerance under a different biological factorization. In the strictly interior model, define $b_k=\lambda_k/\Lambda$, $r_k=s_kd_k$. If only secondary generation or only secondary tolerance varies by antecedent, the exact feasible image requires $0<\eta_{jk}<r_k$; if only secondary establishment or detection varies, it instead requires $0<\eta_{jk}<b_k$. For example, with secondary establishment $s_{k\mid j}$, the rate is $b_k s_{k\mid j} d_k$; existence of a shared $d_k$ is equivalent to $\max(r_k,\max_j\eta_{jk}/b_k)<1$. The other factorizations follow by exchanging the corresponding factors. The two images overlap but neither contains the other at a target whose $b_k$ and $r_k$ are separately identified. With $b_k=1/10,r_k=1/2,\eta_{jk}=1/5$ only the first image is feasible; swapping $b_k,r_k$ gives the reverse. A sink provides only $b_kr_k$, so these projected images coincide. A common antecedent-specific opportunity multiplier instead predicts constant $Q_{jk}$ over the observed targets of that antecedent. A pair-specific unrestricted multiplier need not satisfy this restriction. This gives model-specific falsification tests, not identification of a unique mechanism in the overlap.

The rank calculation is a channel-placement result: cycles and connectivity have no additional identifying effect in independent clades. Secondary marginal overdispersion and primary--secondary covariance give redundant restrictions. For one edge, the quantity $(\operatorname{Var}(Y)-\E[Y])/\E[Y]-\operatorname{Cov}(X,Y)/\E[X]$ vanishes in homogeneous Model B. It also vanishes under common multiplicative exposure, since both terms increase by the same $\lambda\eta\operatorname{Var}(H)$ for $\E[H]=1$. Passing the restriction is not evidence that exposure or within-founder dependence is absent.

\section{Exact intervention-equivalence witness}\label{app:challenge}
A remaining obstacle is that even a perfectly measured baseline trajectory law need not identify the response to an unmeasured challenge. This is familiar in finite-state identifiability and observable-operator theory \cite{blackwell,gilbert,jaeger}. The following exact construction makes the experimental requirement explicit. Let states $1,2,3,\dagger$ emit labels $b,a,a,\dagger$, with initial state 1 and transition matrix
\[
 K=\begin{pmatrix}0&1&0&0\\1/4&1/4&1/4&1/4\\1/2&0&0&1/2\\0&0&0&1\end{pmatrix}.
\]
For $p\in[0,1/2]$, define $S_p$ as the identity except its second row is $(0,1-p,p,0)$. The matrix $K_p=S_p^{-1}KS_p$ remains row-stochastic and nonnegative on this interval (see explicit matrix below). Since $S_p$ preserves the initial vector and ones vector and commutes with emission projectors, the probability of every baseline observation word is unchanged. In contrast, $(K_p)_{13}=p$. A calibrated physical challenge that produces a positive response precisely in state 3 after the first transition has response probability $p$. Thus the same complete baseline law permits at least the interval $[0,1/2]$ of challenge-response probabilities. This is an attainable interval within the displayed family, not a proof of sharpness over all stochastic models. A similarity transformation of the intervention operator as well would preserve the response and merely change coordinates. The independently calibrated physical action is what distinguishes biological hypotheses.

Explicit multiplication gives
\[
 K_p=\begin{pmatrix}
 0&1-p&p&0\\
 (1-2p)/(4(1-p))&1/4&(1+p)/(4(1-p))&(1-2p)/(4(1-p))\\
 1/2&0&0&1/2\\0&0&0&1
 \end{pmatrix}.
\]
Among values for which $S_p$ is invertible, all entries are nonnegative exactly on the declared interval and rows sum to one. If $D_y$ projects onto states emitting label $y$, then $K_pD_y=S_p^{-1}KD_yS_p$ and the product telescopes for every word. The initial row $e_1^\top$ and terminal ones vector are preserved. For example $p=1/3$ yields a first-transition probability $1/3$ into state 3, compared with zero in $K$. The two latent states are not separable by the baseline readout. If their independently calibrated challenge-response probabilities are $a,b$, respectively, the response probability is $a(1-p)+bp$, which identifies $p$ for $a\ne b$. No identification of the challenge calibration follows from baseline equivalence. This witness uses deterministic readout and homogeneous dynamics. It is not a universal identifiability theorem for biological challenges.

\section{A sharp marginal bound on viable fate switching}\label{app:coupling}
Let $a$ and $b$ be the two founder-level distributions over a finite set $\mathcal V$ of mutually exclusive viable destinations and one remaining category $N$. A coupling $\pi$ is a joint distribution with row marginal $a$ and column marginal $b$. Set $I=\sum_{v\in\mathcal V}(b_v-a_v)_+$, where $x_+=\max(x,0)$.
\begin{proposition}
Over all couplings of $a$ and $b$, the minimum of $\sum_{v,w\in\mathcal V:v\ne w}\pi_{vw}$ equals $\max(0,I-a_N)$. If rescue from $N$ into viable destinations is forbidden and a coupling remains feasible, the minimum is $I$.
\end{proposition}
\begin{proof}
A viable destination $v$ with $b_v>a_v$ needs at least $b_v-a_v$ incoming mass from other categories, even when its diagonal mass is maximized. Across all such destinations the total need is $I$. At most $a_N$ can arrive from $N$, proving the lower bound. For attainment retain $\min(a_i,b_i)$ on each viable diagonal. Allocate $\min(a_N,I)$ of the $N$ row to viable deficits, then fill the remaining deficits from viable surplus rows. This is feasible because total viable surplus is $I+a_{\mathcal V}-b_{\mathcal V}=I+b_N-a_N$. If $a_N>I$, no viable switching is needed and remaining $N$ mass and viable surplus fill the $N$ column. If $a_N\le I$, viable surplus fills the remaining $I-a_N$ and the $N$ column. Both yield the lower bound. Without rescue, feasibility requires and is equivalent to $a_N\le b_N$; viable surplus then suffices to fill all $I$, with the excess sent to $N$. \end{proof}
This bound does not identify the joint distribution of counterfactual outcomes or establish a physical switching mechanism. In the presence of additional constraints on feasible couplings it remains a lower bound, but sharpness can change. If categories are simultaneous descendant outcomes rather than exclusive founder fates, the categories must first be defined as mutually exclusive joint outcomes.

\section{Estimation, uncertainty and design diagnostics}\label{app:inference}
\begin{table}[!htb]\centering\small
\caption{Root mean squared error of estimated contrasts, in percentage points, across the alternative seeding regimes. Each entry uses 100 independent two-arm simulations. The measurement designs use different information; the true contrast is $-4$ percentage points.}\label{tab:seedingprecision}
\begin{tabular}{rlrrrrrr}\toprule
Cultures & Method & Poisson & Fixed & Untracked & Matched & Zero-inflated & Operation-only\\
per arm & & & & gamma & gamma & & gamma\\\midrule
2,000 & Intensity & 1.54 & 2.45 & 2.00 & 1.97 & 3.68 & 8.68\\
2,000 & Actual-founder & 1.38 & 1.07 & 1.71 & 1.73 & 2.16 & 1.26\\
2,000 & Known count & 1.26 & 1.09 & 1.44 & 1.43 & 2.05 & 1.29\\
2,000 & Lineage & 0.61 & 0.73 & 0.71 & 0.63 & 0.73 & 0.69\\
10,000 & Intensity & 0.73 & 2.21 & 1.90 & 0.83 & 3.62 & 8.71\\
10,000 & Actual-founder & 0.60 & 0.49 & 0.70 & 0.71 & 1.03 & 0.72\\
10,000 & Known count & 0.53 & 0.45 & 0.62 & 0.66 & 0.92 & 0.64\\
10,000 & Lineage & 0.31 & 0.28 & 0.30 & 0.31 & 0.24 & 0.31\\
\bottomrule\end{tabular}\end{table}

Write $A(s)=\E[Ys^{Y-1}]$, $B(s)=\E[Zs^Y]$, $M_Z=\E[Z]$, $Q=\int_0^1 A(s)/B(s)\,\mathrm{d}s$. At fixed known $m$, the positive-outcome estimator uses $M_ZQ/m$. Its first-order influence function at one observed culture $(Y,Z)$ is
\[
 \frac{(Z-M_Z)Q}{m}+\frac{M_Z}{m}\int_0^1\left[\frac{Ys^{Y-1}}{B(s)}-\frac{A(s)Zs^Y}{B(s)^2}\right]ds.
\]
For matching, with root $t$ and derivative $G'_Z(t)>0$, the influence function is $(Z-M_Z)(1-t)/m-(M_Z/m)(\mathbf{1}_{Y=0}-t^Z)/G'_Z(t)$. Both estimates use sample analogues and standard deviations divided by $\sqrt N$. These formulas presume regularity, stable positive denominators, fixed calibration and independent cultures. They do not include calibration uncertainty. The benchmark uses quadrature consistently for the estimate and its standard error. When exposure influences the mark law, the common exposure factor in the proof no longer cancels: the ratio weights conditional mark derivatives by exposure and by a mark-dependent generating function. The resulting function need not represent the founder-weighted biological target. Calibration sensitivity and alternative observation models should be evaluated before mechanistic interpretation.

For the actual-founder design $p=1-\exp(-J)$ with $J=M_Z Q/m$. Its influence function is $(1-p)$ times the displayed influence function for $J$. For joint estimates use the sum and difference of the individual influence functions before forming their empirical covariance; separate marginal standard errors would ignore their dependence. For binary patterns with two targets, the four entries are $(1-p_{A\cup B},p_{A\cup B}-p_B,p_{A\cup B}-p_A,p_A+p_B-p_{A\cup B})$. Independent arm covariance matrices add for a contrast. The same linear map applied to the covariance matrices gives the pattern-level uncertainty. Point estimates outside the probability simplex are flagged, not interpreted as negative biological frequencies. Nominal intervals are not simultaneous unless adjusted.

If an independent external estimate of $m$ has variance $v_m$, the delta method adds $(p/m)^2v_m$ for the intensity design or $[(1-p)J/m]^2v_m$ for the actual-founder design. These are derivatives of the calibration map, not results of a simulation of calibration error. Shared or paired calibration requires the joint covariance, including sampling/calibration covariance if the same units supply both. Bounds in Appendix~\ref{app:calibration} address systematic uncertainty and must not be confused with sampling intervals. The empirical implementation refuses unstable zero-endpoint denominators and checks quadrature at two resolutions; an exact population limit is not a guarantee of a stable numerical estimate.

For a Poisson founder mean $\lambda$ known in each culture and a binary positive outcome, $\Pp(Y=0)=\exp(-\lambda p)$. Bernoulli information is $\lambda^2/(\exp(\lambda p)-1)$. Differentiation with $x=\lambda p$ gives the unique positive optimum $2(1-e^{-x})=x$ for per-culture information. The per-founder information is proportional to $x/(e^x-1)$ and is decreasing. For independent equal arm sizes, the normal approximation for sample size is $N\ge (z_{.975}+z_{.8})^2[\mathcal I(p_0;\lambda)^{-1}+\mathcal I(p_e;\lambda)^{-1}]/(p_e-p_0)^2$. These calculations assume exact seeding calibration and no clustering. The pilot simulation compares a constant $p$ model against $\log p(\lambda)$ linear in $\log\lambda$, bounded by $p\le1$ over the observed range. The likelihood-ratio test uses the asymptotic one-degree-of-freedom reference distribution. It has limited power and should not be used to establish equivalence by failure to reject.

Under a conditionally Poisson intensity reference with $\E[H]=1$, total variance gives $\operatorname{Var}Z=\E[Z]+c^2\operatorname{Var}H$. This yields the coefficient-of-variation diagnostic in the main text, but no such result follows from a mean restriction alone. For $H=1\pm v$ equiprobably and mean $m$, the homogeneous zero-count estimator has population value $p-\log\cosh(mpv)/m$. The contrast bias is $[\log\cosh(mp_0v)-\log\cosh(mp_ev)]/m$. This is a specific analytic sensitivity example, not a general bias law.

For a multireference diagnostic define $R_i(s)=\E[Z_i s^Y]/\E[Z_i]$. If both references have proportional conditional means for the same seeding variable, $R_1(s)=R_2(s)$. The influence function for $R_i$ is $Z_i(s^Y-R_i)/\E[Z]_i$; the difference must use paired culture-level variability. The implemented fixed-grid diagnostic uses Bonferroni-adjusted normal intervals under independent cultures and finite variance. It is a falsification test only; shared invalid references can pass. Split-sample agreement likewise tests stability, not identification. In particular the recovery integrand is not expected to be constant in $s$; permuting reference counts destroys genuine common exposure and is not a validity test.


\begin{thebibliography}{99}
\bibitem{waddington} Waddington CH. The Strategy of the Genes: A Discussion of Some Aspects of Theoretical Biology. London: George Allen \& Unwin; 1957. \href{https://www.routledge.com/The-Strategy-of-the-Genes/Waddington/p/book/9781315765471}{Publisher reissue record}.
\bibitem{luria} Luria SE, Delbr\"uck M. Mutations of bacteria from virus sensitivity to virus resistance. Genetics. 1943;28:491--511. \href{https://doi.org/10.1093/genetics/28.6.491}{doi:10.1093/genetics/28.6.491}.
\bibitem{rosche} Rosche WA, Foster PL. Determining mutation rates in bacterial populations. Methods. 2000;20:4--17. \href{https://pmc.ncbi.nlm.nih.gov/articles/PMC2932672/}{Full text}.
\bibitem{blount} Blount ZD, Borland CZ, Lenski RE. Historical contingency and the evolution of a key innovation in an experimental population of Escherichia coli. Proc Natl Acad Sci USA. 2008;105:7899--7906. \href{https://doi.org/10.1073/pnas.0803151105}{doi:10.1073/pnas.0803151105}.
\bibitem{schiebinger} Schiebinger G, Shu J, Tabaka M, et al. Optimal-transport analysis of single-cell gene expression identifies developmental trajectories in reprogramming. Cell. 2019;176:928--943.e22. \href{https://pubmed.ncbi.nlm.nih.gov/30712874/}{PMID:30712874}.
\bibitem{weinreb} Weinreb C, Rodriguez-Fraticelli A, Camargo FD, Klein AM. Lineage tracing on transcriptional landscapes links state to fate during differentiation. Science. 2020;367:eaaw3381. \href{https://pubmed.ncbi.nlm.nih.gov/31974159/}{PMID:31974159}.
\bibitem{brouwer} Brouwer AF, Meza R, Eisenberg MC. A systematic approach to determining the identifiability of multistage carcinogenesis models. Risk Anal. 2017;37:1375--1387. \href{https://doi.org/10.1111/risa.12684}{doi:10.1111/risa.12684}.
\bibitem{lu} Lu J, Ding P, Dasgupta T. Treatment effects on ordinal outcomes: causal estimands and sharp bounds. J Educ Behav Stat. 2018;43:540--567. \href{https://doi.org/10.3102/1076998618776435}{doi:10.3102/1076998618776435}.
\bibitem{frangakis} Frangakis CE, Rubin DB. Principal stratification in causal inference. Biometrics. 2002;58:21--29. \href{https://doi.org/10.1111/j.0006-341X.2002.00021.x}{DOI: 10.1111/j.0006-341X.2002.00021.x}.
\bibitem{zhang} Zhang JL, Rubin DB. Estimation of causal effects via principal stratification when some outcomes are truncated by ``death''. J Educ Behav Stat. 2003;28:353--368. \href{https://doi.org/10.3102/10769986028004353}{doi:10.3102/10769986028004353}.
\bibitem{lee} Lee DS. Training, wages, and sample selection: estimating sharp bounds on treatment effects. Rev Econ Stud. 2009;76:1071--1102. \href{https://doi.org/10.1111/j.1467-937X.2009.00536.x}{doi:10.1111/j.1467-937X.2009.00536.x}.
\bibitem{dennis} Dennis EB, Morgan BJT, Ridout MS. Computational aspects of $N$-mixture models. Biometrics. 2015;71:237--246. \href{https://doi.org/10.1111/biom.12246}{doi:10.1111/biom.12246}.
\bibitem{barker} Barker RJ, Schofield MR, Link WA, Sauer JR. On the reliability of $N$-mixture models for count data. Biometrics. 2018;74:369--377. \href{https://doi.org/10.1111/biom.12734}{doi:10.1111/biom.12734}.
\bibitem{elda} Hu Y, Smyth GK. ELDA: extreme limiting dilution analysis for comparing depleted and enriched populations in stem cell and other assays. J Immunol Methods. 2009;347:70--78. \href{https://doi.org/10.1016/j.jim.2009.06.008}{doi:10.1016/j.jim.2009.06.008}.
\bibitem{buchmann} Buchmann B, Gr\"ubel R. Decompounding: an estimation problem for Poisson random sums. Ann Stat. 2003;31:\allowbreak 1054--1074. \href{https://researchportalplus.anu.edu.au/en/publications/decompounding-an-estimation-problem-for-poisson-random-sums/}{Publication record}.
\bibitem{lastpenrose} Last G, Penrose M. Lectures on the Poisson Process. Cambridge University Press; 2017. Chapters 4 and 5. \href{https://www.cambridge.org/core/books/lectures-on-the-poisson-process/97F98804D6A3A378C305C8BE1A0D657D}{Publisher record}.
\bibitem{blackwell} Blackwell D, Koopmans L. On the identifiability problem for functions of finite Markov chains. Ann Math Stat. 1957;28:1011--1015. \href{https://www.jstor.org/stable/2237062}{Publication record}.
\bibitem{gilbert} Gilbert EJ. On the identifiability problem for functions of finite Markov chains. Ann Math Stat. 1959;30:688--697. \href{https://www.jstor.org/stable/2237409}{Publication record}.
\bibitem{jaeger} Jaeger H. Observable operator models for discrete stochastic time series. Neural Comput. 2000;12:1371--1398. \href{https://doi.org/10.1162/089976600300015411}{doi:10.1162/089976600300015411}.
\bibitem{catchpole} Catchpole EA, Morgan BJT. Detecting parameter redundancy. Biometrika. 1997;84:187--196. \href{https://doi.org/10.1093/biomet/84.1.187}{doi:10.1093/biomet/84.1.187}.
\end{thebibliography}
\end{document}